%% file: ANA-TOPQ-2019-02-PAPER.tex
\pdfoutput=1
\documentclass[cernpreprint, atlasdraft=false, UKenglish, texmf, orcidlogo]{atlasdoc}
\usepackage{atlaspackage}
\usepackage{atlasbiblatex}

\usepackage{atlasphysics}
\graphicspath{{logos/}{figures/}}

\usepackage{ANA-TOPQ-2019-02-PAPER-defs}

\usepackage{makecell}
\usepackage{xcolor}
\AtlasTitle{Measurement of $t$-channel production of single top quarks and antiquarks in \(pp\) collisions at 13 TeV using the full ATLAS Run 2 data sample}
\AtlasAbstract{%
The production of single top quarks and top antiquarks via the $t$-channel exchange of a
virtual $W$ boson is measured in proton--proton collisions at a centre-of-mass energy
of 13~TeV at the LHC using $140\,\mathrm{fb^{-1}}$ of ATLAS data.
The total cross-sections are determined to be $\sigma(tq)=137^{+8}_{-8}\,\mathrm{pb}$
and $\sigma(\bar{t}q)=84^{+6}_{-5}\,\mathrm{pb}$ for top-quark and top-antiquark production,
respectively.
The combined cross-section is found to be $\sigma(tq+\bar{t}q)=221^{+13}_{-13}\,\mathrm{pb}$ and the
cross-section ratio is $R_{t}=\sigma(tq)/\sigma(\bar{t}q)=1.636^{+0.036}_{-0.034}$.
The predictions at next-to-next-to-leading-order in quantum chromodynamics are in good agreement
with these measurements.
The predicted value of $R_{t}$ using different sets of parton
distribution functions is compared with the measured value, demonstrating the potential to further
constrain the functions when using this result in global fits.
The measured cross-sections are interpreted in an effective field
theory approach, setting limits at the 95\% confidence level on the strength of a four-quark operator
and an operator coupling the third quark generation to the Higgs boson doublet:
$-0.37 < C_{Qq}^{3,1}/\Lambda^2 < 0.06$ and
$-0.87 < C_{\phi Q}^{3}/\Lambda^2 < 1.42$.
The constraint $|V_{tb}|>0.95$ at the 95\% confidence level is derived from the measured value of
$\sigma(tq+\bar{t}q)$,
assuming that the $Wtb$ interaction is a left-handed weak coupling and
that $|V_{tb}|\gg|V_{td}|,|V_{ts}|$.
In a more general approach, pairs of CKM matrix elements involving top quarks are simultaneously
constrained, leading to confidence contours in the corresponding two-dimensional
parameter spaces.
}

\author{The ATLAS Collaboration}

\AtlasRefCode{TOPQ-2019-02}

\AtlasJournal{JHEP}
\AtlasJournalRef{JHEP 05 (2024) 305}
\AtlasDOI{10.1007/JHEP05(2024)305}

\PreprintIdNumber{CERN-EP-2023-289}

\hypersetup{pdftitle={ATLAS document},pdfauthor={The ATLAS Collaboration}}

\begin{document}

\maketitle

\tableofcontents

\pagebreak

\section{Introduction}

\begin{figure}[h!]
\begin{center}
\subfloat[][]{%
\includegraphics[width=0.43\textwidth]{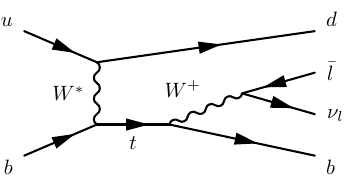}
\label{fig:Feynman_top_quark}
}
\hspace{0.05\textwidth}
\subfloat[][]{%
\includegraphics[width=0.43\textwidth]{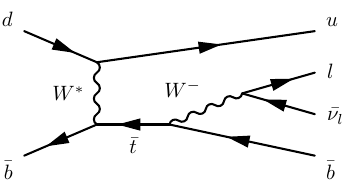}
\label{fig:Feynman_top_antiquark}
}
\caption{\label{fig:feyn_signal}Example Feynman diagrams of
\protect\subref{fig:Feynman_top_quark} single top-quark
and \protect\subref{fig:Feynman_top_antiquark} single top-antiquark production
via the \(t\)-channel exchange of a virtual \wboson at LO in perturbation theory.}
\end{center}
\end{figure}

%
During the 2015--2018 period of operation, known as \RunTwo, the Large Hadron Collider (LHC)
\cite{Evans:2008zzb} provided proton–proton (\(pp\)) collisions at a centre-of-mass energy of
\sqrts, giving the collider experiments access to a previously unexplored kinematic range.
By measuring top-quark production at this energy scale with high precision,
theoretical predictions based on the Standard Model (SM) can be tested and deviations
that might result from energy-dependent non-SM couplings can be searched for.
Top quarks are produced singly in weak charged-current interactions.
The dominant single-top-quark production process at the LHC is characterised by
the \(t\)-channel exchange of a virtual \wboson.
\Fig{\ref{fig:feyn_signal}} depicts example Feynman diagrams of this process
at leading order (LO) in perturbation theory.
A light quark from one of the colliding protons interacts with a \bquark from another proton
by exchanging a virtual \wboson.
Since the valence $u$-quark density of the proton is about twice as high as the valence
$d$-quark density, the production cross-section of single top quarks, \sigt, is
expected to be higher than the cross-section of top-antiquark production, \sigat.

This document presents cross-section measurements of \tq and \tbarq production using the
full data sample recorded with the ATLAS detector~\cite{PERF-2007-01} during \RunTwo of the LHC,
corresponding to an integrated luminosity of 140~\ifb.
Separate measurements of \tq and \tbarq production provide sensitivity to the
parton distribution functions (PDFs) of \(u\)- and \(d\)-quarks~\cite{Nocera:2019wyk},
exploiting the different initial states of the two processes shown in \Fig{\ref{fig:feyn_signal}}.
In addition, the combined cross-section \sigtat and the cross-section ratio
\(\rt=\sigma(tq)/\sigma (\bar t q)\) are measured. The ratio \rt has
better precision than the individual cross-sections
because of partial cancellations of common uncertainties.
The measurements presented here supersede the results obtained by an ATLAS analysis of early \RunTwo data
corresponding to an integrated luminosity of 3.2~\ifb~\cite{TOPQ-2015-15}, significantly improving the precision
due to a larger data sample, better detector calibration, the use of more advanced object
reconstruction~\cite{JETM-2018-05,FTAG-2019-07}, and an improved statistical analysis based on
profile maximum-likelihood fits which fully exploit the statistical power of the data sample.
This analysis also features an improved treatment of systematic uncertainties related to the
modelling of the hard partonic collision and the subsequent hadronisation with event generator programs
based on the Monte Carlo (MC) method. The measurements are compared with fixed-order predictions made at
next-to-next-to-leading-order (\NNLO) in quantum chromodynamics (QCD). The measurement of \rt,
in particular, is compared with predictions based on different PDFs.

The measurements are further interpreted in the context of effective field theory (EFT) to constrain
the Wilson coefficients associated with
the four-quark operator \OqQ and the operator \OphiQ that couples the third quark generation to the
Higgs boson doublet. Both operators potentially contribute to \(t\)-channel single top-quark production
in extensions of the SM. %
Existing limits on the coefficient of \OqQ~\cite{Ethier:2021bye,Ellis:2020unq,Brivio:2019ius}
are based on the combination of published measurements and do not account for reconstruction effects
on EFT signal events, while the analysis presented here employs simulated signal samples and involves template fits to
observed distributions.
The operator \OphiQ features the same Lorentz structure as the \(Wtb\) vertex in the SM and limits on
this operator are thus obtained from the measured value of \sigtat.
Another interpretation sets limits on the coupling strengths of \(Wtq\) vertices, constraining the
products of a left-handed form factor \fl and the absolute values of the Cabibbo--Kobayashi--Maskawa (CKM) matrix elements
\(V_{tb}\), \(V_{ts}\) and \(V_{td}\). The form factor scales the cross-section of \tq production
with \(\fl^2\), but leaves the Lorentz structure of the \(Wtq\) vertices unchanged;
and thus the event kinematics remain unchanged. The SM has \(\fl=1\).

The event selection of the analysis targets \tq and \tbarq events with leptonically
decaying \(W\) bosons.
Consequently, either one isolated electron or muon and high missing transverse momentum are
required. In addition, there must be exactly two hadronic jets with high transverse momentum.
Exactly one of these jets must be identified as originating from a \bquark and is hence
labelled as a \btagged jet, while the second jet is preferentially produced in the
forward direction at high absolute values of pseudorapidity.
The main background processes are top-quark--antiquark (\ttbar) pair production and \wbb production.
Two other single top-quark production processes are also relevant backgrounds: the associated production
of a \wboson and a top quark (\(tW\) production) and the production of $t \bar b$ or $\bar t b$
via the \(s\)-channel exchange of a virtual \wboson. The selected signal and background events are
further separated by constructing discriminants with an artificial neural network (NN).
The output distributions of the NN are used in a maximum-likelihood fit to determine the signal yields and
measure \sigt, \sigat, \sigtat, and \rt.

The CMS Collaboration measured \tq and \tbarq production at \sqrts using a partial \RunTwo
data sample, determining the total cross-sections~\cite{CMS-TOP-17-011},
various differential cross-sections~\cite{CMS-TOP-17-023},
and measuring CKM matrix elements~\cite{CMS-TOP-17-012}.


\section{The ATLAS detector}
%

%
\newcommand{\AtlasCoordFootnote}{%
ATLAS uses a right-handed coordinate system with its origin at the nominal interaction point (IP)
in the centre of the detector and the \(z\)-axis along the beam pipe.
The \(x\)-axis points from the IP to the centre of the LHC ring,
and the \(y\)-axis points upwards.
Cylindrical coordinates \((r,\phi)\) are used in the transverse plane,
\(\phi\) being the azimuthal angle around the \(z\)-axis.
The pseudorapidity is defined in terms of the polar angle \(\theta\) as \(\eta = -\ln \tan(\theta/2)\).
Angular distance is measured in units of \(\Delta R \equiv \sqrt{(\Delta\eta)^{2} + (\Delta\phi)^{2}}\).}

The ATLAS detector covers nearly the entire solid angle around the collision point.\footnote{\AtlasCoordFootnote}
It consists of an inner tracking detector surrounded by a thin superconducting solenoid, electromagnetic and hadronic calorimeters,
and a muon spectrometer incorporating three large superconducting toroidal magnets.

The inner-detector system (ID) is immersed in a \SI{2}{\tesla} axial magnetic field
and provides charged-particle tracking in the range of \(|\eta| < 2.5\).
The high-granularity silicon pixel detector covers the vertex region and typically provides four measurements per track,
the first hit normally being in the insertable B-layer installed before Run~2~\cite{ATLAS-TDR-2010-19,PIX-2018-001}.
It is followed by the silicon microstrip tracker, which usually provides eight measurements per track.
These silicon detectors are complemented by the transition radiation tracker (TRT),
which enables radially extended track reconstruction up to \(|\eta| = 2.0\).
The TRT also provides electron identification information
based on the fraction of hits
above a higher energy-deposit threshold corresponding to transition radiation.

The calorimeter system covers the pseudorapidity range \(|\eta|<4.9\).
Within the region \(|\eta|<3.2\), electromagnetic calorimetry is provided by barrel and
endcap high-granularity lead/liquid-argon (LAr) calorimeters,
with an additional thin LAr presampler covering \(|\eta| < 1.8\)
to correct for energy loss in material upstream of the calorimeters.
Hadronic calorimetry is provided by the steel/scintillator-tile calorimeter,
segmented into three barrel structures within \(|\eta| < 1.7\), and two copper/LAr hadronic endcap calorimeters.
The solid angle coverage is completed with forward copper/LAr and tungsten/LAr calorimeter modules
optimised for electromagnetic and hadronic measurements respectively.

The muon spectrometer (MS) comprises separate trigger and
high-precision tracking chambers measuring the deflection of muons in a magnetic field generated by superconducting air-core toroids.
The field integral of the toroids ranges between \num{2.0} and \num{6.0}~Tm
across most of the detector.
A set of precision chambers covers the region \(|\eta| < 2.7\) with three layers of monitored drift tubes,
complemented by cathode-strip chambers in the forward region, where the background is highest.
The muon trigger system covers the range \(|\eta| < 2.4\) with resistive-plate chambers in the barrel, and thin-gap chambers in the endcap regions.
Interesting events are selected to be recorded by the first-level trigger system implemented in custom hardware,
followed by selections made by algorithms implemented in software in the high-level trigger~\cite{TRIG-2016-01}.
The first-level trigger accepts events from the \SI{40}{\MHz} bunch crossings at a rate below \SI{100}{\kHz},
that the high-level trigger further reduces to record events to disk at about \SI{1}{\kHz}.

An extensive software suite~\cite{ATL-SOFT-PUB-2021-001} is used in the reconstruction and analysis of
real and simulated data, in detector operations, and in the trigger and data acquisition systems of the experiment.


%
\section{Samples of data and simulated events}
%

%
The analysis uses proton--proton ($pp$) collision data recorded with the ATLAS detector in the years 2015 to 2018
at a centre-of-mass energy of 13$\,\TeV$.
After applying data-quality requirements~\cite{DAPR-2018-01}, the data set corresponds to an integrated luminosity
of $140.1\,\ifb$ with a relative uncertainty of $0.83\%$~\cite{DAPR-2021-01}. The LUCID-2 detector~\cite{LUCID-2}
was used for the primary luminosity measurements, complemented by measurements using the inner detector and calorimeters.

Events were selected online during data taking by single-electron or single-muon triggers~\cite{TRIG-2018-05,TRIG-2018-01}.
Multiple triggers were combined in a logical OR to increase the selection efficiency.
The lowest-threshold triggers utilised isolation requirements for reducing the trigger rate.
The isolated-lepton triggers had thresholds in transverse momentum (\pT) of $20\,\GeV$ for muons and $24\,\GeV$ for electrons in 2015 data,
and $26\,\GeV$ for both lepton types in 2016, 2017 and 2018 data.
They were complemented by other triggers with higher \pT thresholds but no isolation requirements
to increase the trigger efficiency.

Sets of simulated events from signal and background processes were produced with
MC-based event generator programs to model the physics processes.
After event generation, the response of the ATLAS detector was simulated using the
\GEANT toolkit~\cite{Agostinelli:2002hh} with a full detector model~\cite{SOFT-2010-01}
or a fast simulation~\cite{ATL-PHYS-PUB-2010-013,AFII} which employed a parameterisation of the calorimeter response.
The fast simulation was used for samples that were employed to evaluate systematic uncertainties associated with the
event generators and for samples used for the EFT interpretation.
To account for additional inelastic \(pp\) collisions in the same and neighbouring bunch crossings (pile-up),
minimum-bias interactions were overlaid on the hard-scattering events at the level of simulated energy depositions.
The minimum-bias events were simulated using \PYTHIA[8.186]~\cite{Sjostrand:2007gs} with the
A3~\cite{ATL-PHYS-PUB-2016-017} set of tuned parameters and the \NNPDF[2.3lo] PDF set~\cite{Ball:2012cx}.
The resulting events were weighted to reproduce the observed pile-up distribution.
The average number of interactions per bunch crossing during the entire data-taking period from 2015 to 2018
is 33.7.

Finally, the simulated events were reconstructed using the same software as applied to the collision data.
The same event selection requirements were applied and the selected events were passed through the same
analysis chain. The multijet background is modelled by dedicated samples of
events selected with slightly modified requirements (see \Sect{\ref{sec:nonprompt_leptons}}).
Corrections are applied to simulated events such that the lepton trigger and reconstruction efficiencies,
jet energy calibration and \btag efficiency are in better agreement with the response observed in data.
More details of the simulated event samples are provided in the following subsections.

\subsection{Simulation of \texorpdfstring{\ttbar}{t\={t}} and single-top-quark production}
\label{sec:samples_top_quark}
Samples of simulated events from \ttbar and single-top-quark production were generated using
the next-to-leading-order (NLO) matrix-element generator
\POWHEGBOX[v2]~\cite{Nason:2004rx,Frixione:2007nw,Frixione:2007vw,Alioli:2009je,Alioli:2010xd,Re:2010bp,Frederix:2012dh},
setting the top-quark mass to \(m_t=172.5\,\GeV\).
For \ttbar and \(tW\) production as well as $s$-channel single-top-quark production ($t\bar{b}$ production)
the \NNPDF[3.0nlo] PDF set~\cite{Ball:2014uwa} was used with the five-flavour scheme.
Following a recommendation given in Ref.~\cite{Frederix:2012dh},
single top-quark production in the \(t\)-channel ($tq$ production) was simulated with the
\NNPDF[3.0nlo\_nf4] PDF set, which implements the four-flavour scheme.
Parton showers, hadronisation, and the underlying event were modelled using \PYTHIA[8.230]
with the A14~\cite{ATL-PHYS-PUB-2014-021} set of tuned parameters and the \NNPDF[2.3lo] PDF set.
The {\POWHEGBOX}+\PYTHIA generator setup applies a matching scheme to the modelling of hard emissions in the two programs.
For \(\ttbar\) production, the matrix-element-to-parton-shower matching is steered by the $\hdamp$ parameter,
that controls the $\pt$ of the first additional gluon emission beyond the \LO Feynman diagram
in the parton shower and therefore regulates the high-\pt emission against which the \ttbar system recoils.
Event generation was run with $\hdamp=1.5\times m_t$~\cite{ATL-PHYS-PUB-2016-020}.
The renormalisation and factorisation scales were set dynamically on an event-by-event basis,
namely to $\muR = \muF = \sqrt{m_t^2 + \pt^2(t)}$ for \ttbar production and to
$\muR = \muF = 4\,\sqrt{m_b^2 + \pt^2(b)}$ for $tq$ production, where $\pt(t)$ is the $\pt$
of the top quark, \(m_b\) is the mass of the \bquark,
and $\pt(b)$ is the $\pt$ of the $b$-quark originating from the initial-state gluon that
splits into a $\bbbar$ pair. The scale choice for $tq$ production follows a recommendation in
Ref.~\cite{Frederix:2012dh}.
When generating $tW$ events, the scales were set to \(\muR = \muF = m_t\) and
the diagram-removal scheme~\cite{Frixione:2008yi} was employed to treat the interference with \ttbar\ production~\cite{ATL-PHYS-PUB-2016-020}.

In the case of \(tq\) production, top-quark decays were modelled by \MADSPIN~\cite{Frixione:2007zp,Artoisenet:2012st},
while in the case of
$\ttbar$, \(t\bbar\) and \(tW\) production top-quark decays were handled by \POWHEGBOX directly.
The decays of bottom and charm hadrons were simulated using the \textsc{EvtGen~1.6.0} program~\cite{Lange:2001uf}.

The sample of simulated \ttbar events was normalised to a total cross-section of \(\sigma(\ttbar)=834\pm33\,\)pb
(relative uncertainty: \(4.0\%\)),
the value obtained from \NNLO predictions from the \textsc{Top++~2.0}
program (see Ref.~\cite{Czakon:2011xx} and references therein),
which includes the resummation of next-to-next-to-leading logarithmic (NNLL) soft-gluon terms.
The predicted cross-sections of \(tq\) and \(\tbar q\) production used to normalise the corresponding samples
of simulated events are \(\sigt=134.2\pm2.2\,\)pb and \(\sigma(\bar{t}q)=80.0\pm1.6\,\)pb
(relative uncertainties: \(1.6\%\) and \(2.0\%\), respectively) and
were calculated with the \MCFM[10.1] program~\cite{Campbell:2020fhf} at \NNLO in QCD.
The quoted uncertainties include the uncertainties related to a variation of \(\muR\) and \(\muF\),
the uncertainty in the PDFs and in the value of the strong coupling constant \alphas.
The scale uncertainty is determined by varying \(\muR\) and \(\muF\) independently up and down
by a factor of two, whilst never allowing them to differ by a factor greater than two from each other.
The combined PDF and \alphas uncertainties were determined at the 68\% confidence level (CL) according to the Hessian
representation of the PDF4LHC21 PDF set~\cite{PDF4LHCWorkingGroup:2022cjn}.
The total cross-section for $t\bar{b}$ production was computed at NLO in QCD with
the \HATHOR[2.1] program~\cite{Aliev:2010zk,Kant:2014oha} and the corresponding sample
of simulated events was normalised to \(\sigma(t\bbar+\tbar b)=10.32\pm0.38\,\)pb
(relative uncertainty: \(3.7\%\)).
The cross-section used for normalising the \(tW\) sample is
\(\sigma(tW+\tbar W)=79.3\pm2.9\,\)pb (relative uncertainty: \(3.7\%\))~\cite{Kidonakis:2021vob}.
All cross-section calculations assume \(m_t=172.5\,\GeV\).

\subsection{Simulation of $W$+jets, $Z$+jets and diboson production}
The production of $W$ and $Z$ bosons in association with jets, including heavy-flavour jets,
was simulated with the \SHERPA[2.2.1] generator~\cite{Bothmann:2019yzt}. In this setup,
NLO-accurate matrix elements for up to two partons and LO-accurate matrix elements for up to four partons
are calculated with the \textsc{Comix}~\cite{Gleisberg:2008fv} and
\OPENLOOPS[1]~\cite{Buccioni:2019sur,Cascioli:2011va,Denner:2016kdg} libraries.
The default \SHERPA parton shower~\cite{Schumann:2007mg} based on Catani--Seymour dipole factorisation and the
cluster hadronisation model~\cite{Winter:2003tt} were used. The generation employed the dedicated set of tuned
parameters developed by the \SHERPA authors and the \NNPDF[3.0nlo] PDF set.

The NLO matrix elements of a given jet multiplicity are matched to the parton shower using a colour-exact
variant of the MC@NLO algorithm~\cite{Hoeche:2011fd}. Different jet multiplicities are then merged into an
inclusive sample using an improved CKKW matching procedure~\cite{Catani:2001cc,Hoeche:2009rj} that is extended
to NLO accuracy using the MEPS@NLO prescription~\cite{Hoeche:2012yf}.
The merging threshold was set to $20\,\GeV$. The $W$+jets and $Z$+jets samples are normalised to NNLO
predictions~\cite{Anastasiou:2003ds} of the total cross-sections, obtained with the
\textsc{FEWZ} package~\cite{Gavin:2010az}.

After event generation and before detector simulation, the \(W\)+jets and \(Z\)+jets samples were subjected
to hadron-flavour filters. Events in which at least one \(b\)-hadron is present were selected and form \(b\)-filtered
samples.
The production of a \(W\) boson in association with \(b\)-hadrons is dominated by processes in which a radiated
high-\pt gluon splits into a \(b\bbar\) pair. This class of background processes is thus called \wbb production.
Samples with an applied \(c\)-filter were produced by vetoing events that pass the \(b\)-filter described
above and requiring at least one \(c\)-hadron to be present. Two different classes of physics
processes contribute to the \(c\)-filtered samples, flavour production via gluon splitting leading
to \(W\)+\(c\bar{c}\) and a second class of processes with a down-type quark and a gluon in the initial
state, leading to the production of a single \(c\)-quark in the final state via \(qg\rightarrow W+c\).
To represent both classes of processes the associated production of a \(W\) boson and \(c\)-jets is
denoted as \(W+c(\bar{c})\) production. Generated events of $W$+jets production that remain
after applying the \(b\)-filter and the \(c\)-filter as a veto constitute \(W\)+light-quark-jet production.
The contribution of this process to the expected event yields is much smaller than
the contributions of \wbb and \(W+c(\bar{c})\) production due to the tight \btag requirement made, and
therefore the \(W\)+light-quark-jet contribution is merged with the contribution of the \wbb process in the statistical analysis.
The \(Z\)+jets samples are treated with the same hadron-flavour filtering scheme as the $W$+jets samples,
leading to \(b\)-filtered, \(c\)-filtered and light-flavour samples. However, $Z$+jets production
is a minor background in the analysis and therefore the flavour split is not used in the statistical analysis.

Samples of on-shell diboson production ($WW$, $WZ$ and $ZZ$) were simulated with the same
\SHERPA setup as described above for \Wjets and \Zjets production.
The matrix elements considered contain all diagrams with four electroweak vertices and were calculated at
NLO accuracy in QCD for up to one additional parton and at LO accuracy for up to three additional parton emissions.
The diboson event samples are normalised to the total cross-sections provided by \SHERPA.

\subsection{Simulation and modelling of multijet production}
Events featuring generic high-\(\pt\) multijet production may satisfy the event selection if a jet is misidentified
as an electron or muon, or if real electrons or muons coming from hadron decays inside the jets satisfy the
isolation requirements. The former are called \textit{fake leptons}, the latter \textit{non-prompt leptons}.
In addition, non-prompt electrons occur as a result of photon conversions in the detector material.
Multijet events with fake electrons or non-prompt electrons are modelled with a sample of simulated dijet events,
while events with non-prompt muons are modelled with collision data,
described in~\Sect{\ref{sec:nonprompt_leptons}}.
The number of events with fake muons is negligible.
The dijet event sample was generated using \PYTHIA[8.186] with LO matrix elements for dijet production and
interfaced to a \pt-ordered parton shower. The scales \(\muR\) and \(\muF\) were set to the square root of the
geometric mean of the squared transverse masses of the two outgoing particles in the matrix element,
\(\muR = \muF = \sqrt[4]{(p_{\mathrm{T},1}^2 + m_1^2) (p_{\mathrm{T},2}^2 +  m_2^2)}\).
At generator level, a filter was applied that required the existence of one jet with
$\pt > 17\,\GeV$, which was formed by running a jet clustering algorithm over the stable particles of the generated
events. The generation used the \NNPDF[2.3lo] PDF set and the A14 set of tuned parameters.
The generated sample of dijet events is used to model the kinematics of electron events of the
multijet background when producing template distributions,
while the rate of the multijet background is estimated in a data-driven way using dedicated control
regions (CRs) described in \Sect{\ref{sec:cr_definition}}.

\subsection{Samples for the EFT and CKM interpretations}
\label{sec:samples_EFT_CKM}
For interpreting the measurement in the framework of EFT, samples of \tq (\(t\)-channel)
and $t\bbar$ (\(s\)-channel) production were generated
with \MGNLO[2.7.3] using the SMEFTatNLO-NLO model~\cite{Degrande:2020evl} with the five-flavour scheme
and the \NNPDF[3.0nlo] PDF set. The operator \(\OqQ\) was activated, which introduces
a four-quark contact interaction. Separate samples of simulated events were generated for each
\(\CqQ\in\left\{-0.6,\,-0.2,\,0.0,\,0.4,\,1.0\right\}\)
for single top-quark and top-antiquark production.
The setting \(\CqQ=0.0\) corresponds to the SM.
Each sample includes both \(tq\) and \(t\bbar\) production.
The SM production of the two processes is covered as well as the production via the four-quark
operator \(\OqQ\), and the interference of SM and non-SM amplitudes.
The generated events were showered with \PYTHIA[8.244] using the A14 set of tuned parameters and
the \NNPDF[2.3lo] PDF set.
In these samples, the top-quark is assumed to decay to \(W^+b\) with a branching ratio of 100\%.

The generalised CKM interpretation is based on samples of \tq and \(\tbar q\) events generated with \MGNLO[2.9.9] using the
\NNPDF[3.0nlo] PDF set. Eight samples were generated in which all different combinations of \(Wtq\) vertices with
\(q\in\{d, s, b\} \) are considered for the production and the decay vertex,
except for the dominant mode that has a \(Wtb\) vertex on the production and the decay side.
The four-flavour scheme was used for both samples in which the top quark originates from a $b$-quark.
The other six samples were generated based on the five-flavour scheme. Parton showers were simulated with
\PYTHIA[8.307] using the A14 set of tuned parameters and the \NNPDF[2.3lo] PDF set.
The decay of top quarks was simulated with \MADSPIN preserving all spin correlations, while
\(W\) bosons coming from the top-quark decays were forced to decay leptonically.
The samples are normalised to cross-sections calculated with \MGNLO assuming that the CKM matrix elements
involved are equal to one, and were simulated with the full detector simulation.

The main background, \ttbar production, also involves \(Wtq\) vertices when the top quark and antiquark decay.
To facilitate a consistent treatment of the \ttbar background eight additional samples were generated with
\POWHEGBOX[v2], implementing all combinations of \(Wtq\) decay vertices, except for the nominal channel
that involves two \(Wtb\) vertices. For the alternative samples, the parton shower was simulated with
\PYTHIA[8.307] using the A14 set of tuned parameters and the \NNPDF[2.3lo] PDF set.
The top-quark decay was handled with \MADSPIN.
%
%


\section{Object reconstruction and event selection}
%


The partonic final state of the \(tq\) signal process comprises a charged lepton, a neutrino, a \bquark
and a light quark (see \Fig{\ref{fig:feyn_signal}}) and is reconstructed by identifying corresponding objects
measured in the detector, such as electron and muon candidates, and hadronic jets.
The presence of a high-$\pt$ neutrino is indicated by large missing transverse momentum.

\subsection{Object definitions}
Events are required to have at least one vertex reconstructed from at least two ID tracks with transverse momenta
of $\pT>0.5\,\GeV$. The primary vertex of an event is defined as the vertex with the highest sum of $\pT^2$ over
all associated ID tracks~\cite{ATL-PHYS-PUB-2015-026}.

Electron candidates are reconstructed by matching a track in the ID to clusters of energy deposits in
the electromagnetic calorimeter~\cite{EGAM-2018-01}.
The pseudorapidity of clusters, $\eta_{\mathrm{cluster}}$, is required to be in the range of
$|\eta_{\mathrm{cluster}}|<2.47$.
However, clusters are excluded if they are in the transition region $1.37<|\eta_{\mathrm{cluster}}|<1.52$
between the barrel and endcap electromagnetic calorimeters.
Electron candidates must have $\pt>10\,\GeV$.
A likelihood-based discriminant is constructed to simultaneously evaluate several properties of
electron candidates, including shower shapes in the electromagnetic calorimeter, track quality, and
the detection of transition radiation produced in the TRT. By placing a requirement on the discriminant, the
selection of true electrons is enhanced, while photon conversions and hadrons misidentified as electrons
are largely rejected.
Two categories of electrons with different identification quality are defined~\cite{EGAM-2018-01}: the first category implements
\textit{Tight} identification criteria and features a high rejection of non-prompt or fake electrons,
while the second category with \textit{Loose} identification criteria has higher efficiency at the price of lower purity in
prompt electrons.
Electrons from decays of weak gauge bosons with \(\pt(e)>15\,\GeV\) satisfy the \textit{Tight} (\textit{Loose})
criteria with an average efficiency of 80\% (93\%).

Muon candidates are reconstructed by combining tracks in the MS with tracks in the ID~\cite{MUON-2018-03}.
The tracks must be in the range of $|\eta|<2.5$ and have $\pt>10\,\GeV$.
Similarly to electrons, two levels of identification criteria are applied, defining \textit{Medium} and
\textit{Loose} quality categories of muon candidates~\cite{MUON-2018-03}. Muons originating from $W$ bosons in $t\tbar$ events
with $\pt(\mu)>10\,\GeV$ satisfy the \textit{Medium} (\textit{Loose}) quality criteria with an efficiency of
97\% (99\%).

The tracks matched to electron and muon candidates must point to the primary vertex, which is ensured
by requirements imposed on the transverse impact-parameter significance,
$|d_{0}/\sigma(d_{0})|<5$ for electrons and $|d_{0}/\sigma(d_{0})|<3$ for muons,
and on the longitudinal impact parameter, $\Delta{}z_0$, for which
$|\Delta{}z_{0}\sin(\theta)|<0.5$~mm must be satisfied for both of the lepton flavours.
Non-prompt and fake leptons are efficiently rejected using multivariate discriminants~\cite{MUON-2018-03} computed with
boosted decision trees that combine electromagnetic shower shapes, track information from
the ID, and a discriminant used to identify \bjets.
Prompt muons with a \pt between \(20\) and \(100\,\GeV\) satisfy the imposed isolation requirement with an
efficiency of 87\%, while the efficiency for muons from semileptonic decays of bottom or charm hadrons is 0.5\%.
Scale factors are used to correct the efficiencies in simulation to match the efficiencies
measured for the electron~\cite{TRIG-2018-05} and muon~\cite{TRIG-2018-01} triggers,
and the reconstruction, identification and isolation criteria~\cite{EGAM-2018-01,MUON-2018-03}.

Jets are reconstructed from particle-flow objects~\cite{PERF-2015-09} with the \antikt\ clustering
algorithm~\cite{Cacciari:2008gp,Fastjet} using a radius parameter of 0.4.
This algorithm matches topological clusters~\cite{PERF-2014-07} in the calorimeters to selected
tracks in the ID. The energy of tracks is subtracted from the matched topological clusters and
both the tracks and the energy-subtracted topological clusters are used as input to the clustering.
The jet energy is calibrated by applying several simulation-based corrections and techniques
correcting for differences between simulation and data~\cite{JETM-2018-05}.
The jets must fulfil $\pt > \SI{30}{\GeV}$ and $|\eta|<4.5$.

To suppress jets originating from pile-up collisions, several track-based variables are combined
with a multivariate technique in the jet-vertex-tagger~(JVT) discriminant~\cite{PERF-2014-03}.
Jets with $\pt < \SI{60}{\GeV}$ and $|\eta|<2.4$ are required to have a JVT-discriminant above $0.5$,
which corresponds to an efficiency of $92\%$ for non-pile-up jets,
while \SI{98}{\%} of jets from pile-up events are rejected.
For jets with \(\pt < 60~\GeV\) and \(|\eta|>2.5\), the forward-jet-vertex-tagger~(fJVT)~\cite{PERF-2016-06}
is used and an fJVT value below \(0.4\) is required, resulting in efficiencies of 85\% for
hard-scattering jets and 50\% for pile-up jets. In addition, the jet must satisfy a timing condition.
Differences in the efficiencies of the JVT and fJVT requirements between collision data and
simulation are corrected by corresponding scale factors.

Jets containing \(b\)-hadrons are identified (\(b\)-tagged) with the
\texttt{DL1r} algorithm,
which uses a deep feed-forward neural network with several \(b\)-tagging algorithms
as inputs~\cite{FTAG-2019-07}.
These input algorithms exploit the impact parameters of charged-particle tracks, the properties of reconstructed
secondary vertices and the topology of \(b\)- and \(c\)-hadron decays inside the jets.
The requirement on the \texttt{DL1r} discriminant is chosen such that the efficiency of tagging \bjets
with \(\pt>\SI{20}{\GeV}\) produced
in simulated dileptonic \(t \bar t\) events is \(\SI{60}{\percent}\).
Differences in the \(b\)-tagging efficiency between collision data and simulation are corrected with
simulation-to-data scale factors derived from \(t\bar{t}\) events. The scale factors are determined
as a function of jet \(\pt\) and are found to be consistent with unity within uncertainties.
The obtained scale factors depend on the parton-shower generator used to produce the \(\ttbar\) samples.
When using samples produced with a different parton-shower generator,
for example \SHERPA, to model \wjets events, or when evaluating systematic uncertainties with a setup
based on \HERWIG, additional correction factors called MC-to-MC scale factors are applied.
Since the \texttt{DL1r} algorithm uses measurements from the ID,
the identification of \bjets is limited to the region with \(\abseta < 2.5\).

To avoid double-counting objects satisfying more than one selection criterion,
a procedure called \textit{overlap removal} is applied. Reconstructed objects defined with
\textit{Loose} quality criteria are removed in the following order:
electrons sharing an ID track with a muon;
jets within \(\Delta R = 0.2\) of an electron, thereby avoiding double-counting electron energy deposits as jets;
electrons within \(\Delta R = 0.4\) of a remaining jet, for reducing the impact of non-prompt electrons;
jets within \(\Delta R = 0.2\) of a muon if they have two or fewer associated tracks with $\pT > \SI{0.5}{\GeV}$;
and muons within \(\Delta R = 0.4\) of a remaining jet, reducing the rate of non-prompt muons.

The missing transverse momentum \vecpmiss is reconstructed as the negative vector sum of the $\pt$ of
the reconstructed leptons and jets, as well as ID tracks that point to the primary vertex
but are not associated with a reconstructed object~\cite{PERF-2016-07}.
The latter contribution to \vecpmiss is named soft-track component.
The magnitude of \vecpmiss is denoted by \met.

\subsection{Modelling of non-prompt and fake leptons}
\label{sec:nonprompt_leptons}
Events of the multijet background with an identified electron candidate are modelled using the \textit{jet-electron}
method~\cite{ATLAS-CONF-2014-058}. Simulated events from dijet production
are selected if they contain a jet depositing a large fraction (${>}80\%$) of its energy
in the electromagnetic calorimeter.
This jet is classified as an electron, labelled as the jet-electron, and is treated in the
subsequent steps of the analysis in the same way as a properly identified prompt electron as defined in the previous section.
The jet-electrons must satisfy the nominal \pT and \(|\eta|\) requirements,
but electron identification requirements are not applied.

Multijet events with non-prompt muons are modelled with collision events highly enriched in
non-prompt muons~\cite{ATLAS-CONF-2014-058}. Starting from the same sample of collision events
as the nominal selection, a subset of events enriched in non-prompt muons is obtained by inverting
or modifying some of the muon isolation requirements, such that the resulting sample does not overlap
with the nominal sample.
The kinematic requirements on $\pt$ and $|\eta|$ are the same as for the nominal muon selection.


\renewcommand\arraystretch{1.7}

\subsection{Event selection and definition of signal regions}
Candidate events are required to have exactly one charged lepton ($\ell$) with $\pt(\ell)>28\,\GeV$,
either an electron of \textit{Tight} quality or a muon of \textit{Medium} quality.
The charged lepton is required to match the object that caused the event to pass
a single-lepton trigger.
To reduce contributions from $\ttbar$ events in the dilepton decay channel, any event with an
additional lepton satisfying the \textit{Loose} quality conditions with
$\pT > \SI{10}{\gev}$ is rejected.

Multijet events containing fake or non-prompt leptons tend to have low \met and
low $W$ transverse mass, in contrast to events with prompt leptons from $W$ and $Z$ decays.
The $W$ transverse mass is defined as
\begin{equation}
\mtw = \sqrt{2 \pT(\ell) \MET \left(1-\cos \Delta \phi \left(\vecpmiss, \ell \right) \right)}\,,
\nonumber
\end{equation}
using the difference between the azimuthal angles of \(\vecpmiss\) and the charged lepton,
\(\Delta\phi \left(\vecpmiss, \ell \right)\).
To reduce the multijet background, $\met > 30\,\GeV$ and $\mtw > \SI{50}{\gev}$ are applied
as selection requirements.

Exactly two jets with \(\pT >30\,\GeV\) and \(|\eta|<4.5\) are required.
Exactly one of these jets must be \btagged, while the second jet must fail to meet the \btag requirement.
The latter jet is therefore called the \textit{untagged} jet.
The \btagged jet is explicitly required to have \(|\eta|<2.5\).
Events with forward jets with \(2.3<|\eta|<4.5\) are removed if at least one of the jets
has \(\SI{30}{\gev} < \pT < \SI{35}{\gev}\),
leading to an improved modelling of the \(|\eta|\) distribution of untagged jets in the given regime.

To further suppress the multijet background and to remove poorly reconstructed leptons
with low \(\pt\), an additional requirement is applied based on the azimuthal angle
between the charged lepton and the leading jet ($j_1$), i.e.\ the jet with the largest \(\pt\).
This quantity is denoted by \(\Delta \phi \left(j_1,\ell \right)\). The imposed requirement is
\begin{equation}
\label{eq:triangular_cut}
\pT\left(\ell\right) > \SI{40}{\gev} \cdot \frac{|\Delta \phi\left(j_1, \ell \right)|}{\pi},
\end{equation}
which leads to a tighter \pT requirement on the charged lepton than the baseline definition
if the leading jet and the charged lepton
have a back-to-back topology, namely if \(|\Delta \phi(j_1,\ell)|>0.7\pi\).
For the maximum separation \(|\Delta \phi(j_1,\ell)|=\pi\) between the two objects,
\(\pT(\ell)>\SI{40}{\gev}\) must be satisfied.

Furthermore, an additional selection criterion is imposed on the invariant mass of the charged lepton and the $b$-tagged jet,
$m(\ell b)$. Since the off-shell region for top-quark decays is not included in the calculation
of the matrix element of the event generator, it is not modelled well.
Therefore, the tail of the \(m(\ell b)\) distribution is removed by requiring
\(m(\ell b)< \SI{160}{\gev}\); this imposes a threshold that is slightly above the kinematic
limit at LO, $m(\ell b)^2_\text{limit} = m^2_t - m^2_W$.

Two separate signal regions (SRs) are defined for events with a positively or a negatively charged lepton.
These regions are denoted \srp and SR \texttt{minus}, respectively.

\subsection{Control regions for the multijet background}
\label{sec:cr_definition}
Since the misidentification of jets as electrons or muons are not well modelled
by the detector simulation, the rate of the multijet background is determined in a
data-driven way by including dedicated CRs in the fits of the statistical analysis.
The rate of fake and non-prompt electrons is constrained in four CRs that are defined by
the same selection criteria as the two SRs but inverting the \(\met\) requirement.
Since the relative numbers of electrons detected in the barrel ($|\eta|<1.37$) and
endcap ($|\eta|>1.52$) sections of the electromagnetic calorimeter are not modelled
well enough by the sample of simulated dijet events, separate CRs are defined for the barrel and
endcap regions and are denoted CR \texttt{B-e-plus}, CR \texttt{B-e-minus},
CR \texttt{EC-e-plus} and CR \texttt{EC-e-minus}.
Only the event yields in these regions are included in the maximum-likelihood fit.
The rate of non-prompt muons is constrained in two CRs defined by the same
selection criteria as used for the two SRs but inverting the requirement on the \(\pt\) in
\Eqn{\ref{eq:triangular_cut}}.
The two CRs are named \crmup and \crmum. The distributions of the difference
in the azimuthal angles of \(\vecpmiss\) and the muon, \(\Delta\phi(\vecpmiss, \mu)\), are
included in the maximum-likelihood fits.
\Tab{\ref{tab:CRdef}} provides a summary of the definition of the CRs.
\begin{table}[!t]
\begin{center}
\begin{tabular}{ll}
\toprule
CR name & Requirement \\
\midrule
\texttt{B-e-plus}   & \(q_e/e=+1\), \(|\eta(e)|<1.37\), \(\met<30\,\GeV\) \\
\texttt{B-e-minus}  & \(q_e/e=-1\), \(|\eta(e)|<1.37\), \(\met<30\,\GeV\) \\
\texttt{EC-e-plus}  & \(q_e/e=+1\), \(|\eta(e)|>1.52\), \(\met<30\,\GeV\) \\
\texttt{EC-e-minus} & \(q_e/e=-1\), \(|\eta(e)|>1.52\), \(\met<30\,\GeV\) \\
\crmup & \(q_\mu/e=+1\),
\(28\,\GeV<\pT\left(\mu\right) < \SI{40}{\gev} \cdot \frac{|\Delta \phi\left(j_1, \ell \right)|}{\pi}\)\\
\crmum & \(q_\mu/e=-1\), \(28\,\GeV<\pT\left(\mu\right) < \SI{40}{\gev} \cdot \frac{|\Delta \phi\left(j_1, \ell \right)|}{\pi}\) \\
\bottomrule
\end{tabular}
\vspace*{2mm}
\caption{\label{tab:CRdef}Summary of the definition of the CRs.}
\end{center}
\end{table}
%


\section{Separation of signal from background events}
%

%
An artificial neural network is used to separate signal and background events in the two SRs
by combining several kinematic variables into an optimised NN discriminant named \dnn.
In addition to variables derived from the reconstructed objects, the NN builds on a
reconstruction of the \(W\) boson and the top quark.
The reconstruction of the leptonically decaying \(W\) boson requires the determination of the neutrino momentum.
While the \(x\)- and \(y\)-components of the neutrino momentum, \(p_x(\nu)\) and \(p_y(\nu)\), are approximated
by the components of \vecpmiss, the \(z\)-component, \(p_z(\nu)\), is determined by constraining the mass of
the reconstructed \(W\) boson to match the measured world average~\cite{Workman:2022ynf}.
If the resulting quadratic equation has two real solutions, the one with the smallest \(|p_z(\nu)|\) is chosen.
In the case of complex solutions, which occur due to the limited \(\MET\) resolution, a kinematic fit is
performed that rescales the \(p_x(\nu)\) and \(p_y(\nu)\) such that the imaginary part vanishes and
at the same time the distance between the transverse components of the neutrino momentum and \vecpmiss is
minimised~\cite{Chwalek:2010vka}. The \(W\) boson is formed by adding the four-vectors of the reconstructed neutrino and
the charged lepton. The top quark is reconstructed by adding the four-vector of the \(W\) boson and the
\(b\)-jet.

The NN is implemented using the NeuroBayes package~\cite{feindt-2004,Feindt:2006pm}, which combines
a three-layer feed-forward NN with a complex and robust preprocessing of the input variables before they
are presented to the NN. The preprocessing produces a ranking of the input variables based on an algorithm
employing the total correlation of a set of variables to the target function, which assumes the
value \(1\) for signal and \(0\) for background events~\cite{TOPQ-2011-14}.
Utilising this ranking, NNs with different numbers of variables
are trained, the full analysis is performed and the expected uncertainty of the measurement is determined.
Networks using more input variables tend to result in measurements with lower uncertainties in \sigt, \sigat,
\sigtat, and \rt. However, when employing 15--30 variables,
only marginal further improvements are found if more variables are added.
As a result, the 17 highest-ranking input variables
are chosen for training the final NN. These input variables are listed and described in
Table~\ref{tab:NNinputVars}.
\begin{table}[!t]
\begin{center}
\begin{tabular}{rll}
\toprule
No. & Symbol & Description \\
\midrule
1. & \(m(jb)\)                            & Invariant mass of the untagged jet (\(j\)) and the \(b\)-tagged jet (\(b\)) \\
2. & \(|\eta(j)|\)                        & Absolute value of the pseudorapidity of the untagged jet \\
3. & \(|\Delta\pt(W, jb)|\)               & \makecell[l]{Absolute value of the difference in transverse momentum between the \\
reconstructed \(W\) boson and the jet pair} \\
4. & \(|\Delta\phi(W, jb)|\)              & \makecell[l]{Absolute value of the difference in azimuthal angle between the recon- \\
structed \(W\) boson and the jet pair}\\
5. & \(m(t)\)                             & Invariant mass of the reconstructed top quark\\
6. & \(|\Delta\eta(\ell ,j)|\)            & \makecell[l]{Absolute value of the difference in pseudorapidity between the charged \\
lepton (\(\ell\)) and the untagged jet} \\
7. & \(\Delta R(\ell ,j)\)                & Angular distance of the charged lepton and the untagged jet \\

8. & \(|\Delta\eta(b,\ell )|\)            & \makecell[l]{Absolute value of the difference in pseudorapidity between the \(b\)-tagged jet \\
and the charged lepton}\\
9. & \(\mtw\)                             & Transverse mass of the \(W\) boson\\
10. & \(m(\ell b)\)                       & Invariant mass of the charged lepton and the \(b\)-tagged jet\\
11. & \(\HT(\ell ,\mathrm{jets}, \met)\)  & \makecell[l]{Scalar sum of the transverse momenta of the charged lepton and the \\
jets and \(\MET\)} \\
12. & \(|\Delta\eta(b,j)|\)               & Absolute value of the difference in the pseudorapidity of the two jets\\
13. & \(|\Delta\phi(j, t)|\)              & \makecell[l]{Absolute value of the difference in the azimuthal angle between the \\
untagged jet and the reconstructed top quark} \\
14. & \(\cos\theta^*(\ell ,j)\)           & \makecell[l]{Cosine of the angle \(\theta^*\) between the charged lepton and the untagged \\
jet in the rest frame of the reconstructed top quark} \\
15. & \(|\eta(\ell)|\)                    & Absolute value of the pseudorapidity of the charged lepton \\

16. & \(S\)                               & \makecell[l]{Sphericity defined as the sum of the 2nd and 3rd largest eigenvalues\\
of the sphericity tensor multiplied by \(3/2\)} \\
17. & \(|\Delta \pT (\ell ,j) |\)         & \makecell[l]{Absolute value of the difference in transverse momentum of the charged \\
lepton and the untagged jet} \\
\bottomrule
\end{tabular}
\vspace*{2mm}
\caption{\label{tab:NNinputVars}The 17 variables used for the training of the NN ordered by their discriminating
power.
The sphericity tensor \(S^{\alpha\beta}\) used to define the sphericity \(S\) is formed with the three-momenta \(\vec{p}_i\)
of the reconstructed objects,
namely the jets, the charged lepton and the reconstructed neutrino. The tensor is given by
\(S^{\alpha\beta}=\frac{\sum_i p_i^\alpha p_i^\beta}{\sum_i |\vec{p}_i|^2}\) where \(\alpha\) and \(\beta\) correspond to
the spatial components $x$, $y$ and $z$.}
\end{center}
\end{table}
The probability densities of the two most discriminating variables, \(m(jb)\) and \(|\eta(j)|\), are shown for
the \tq signal process, and the \(t\tbar\) and the \wbb backgrounds in \Fig{\ref{fig:input_vars_shapes}} for \srp.
The symbol \(j\) represents the untagged jet.
\begin{figure}[tb!]
\centering
\subfloat[][]{%
\includegraphics[width=0.46\textwidth]{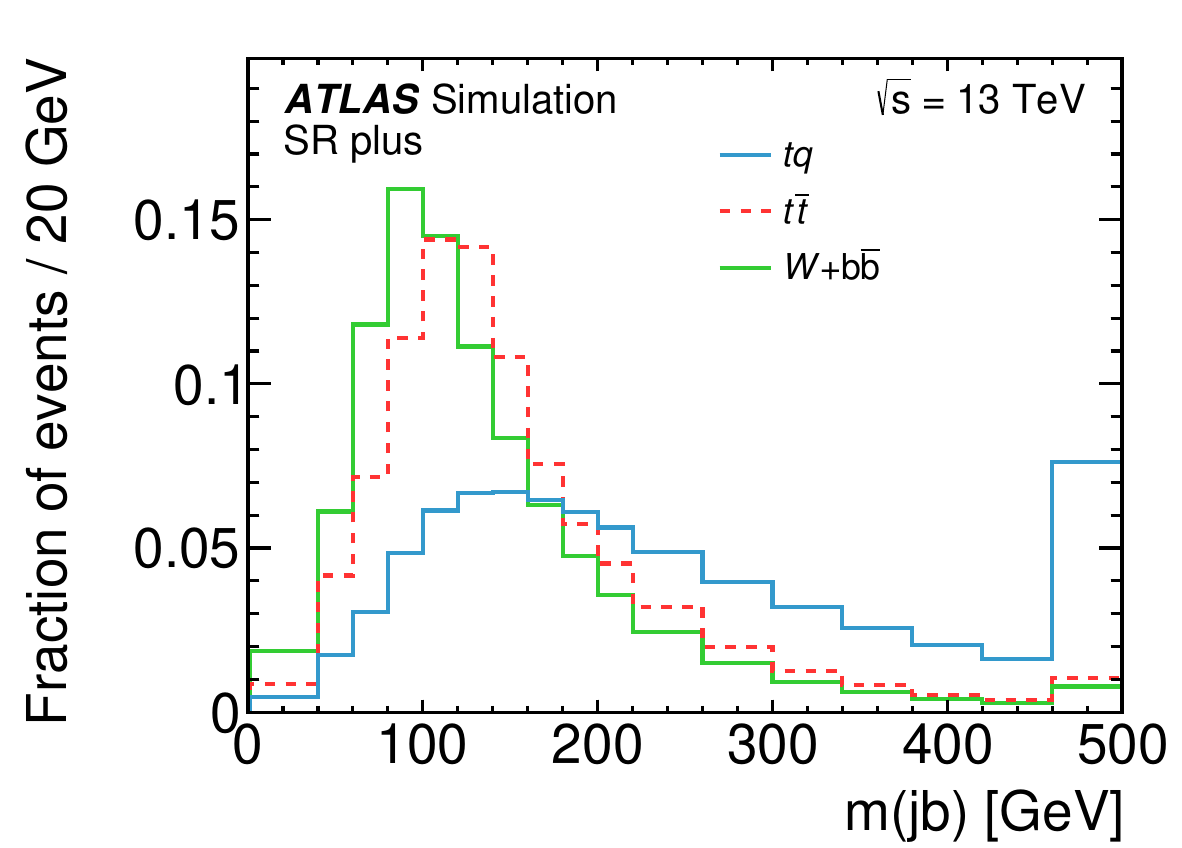}
\label{fig:input_var_shapes_mjb}
}
\quad
\subfloat[][]{%
\includegraphics[width=0.46\textwidth]{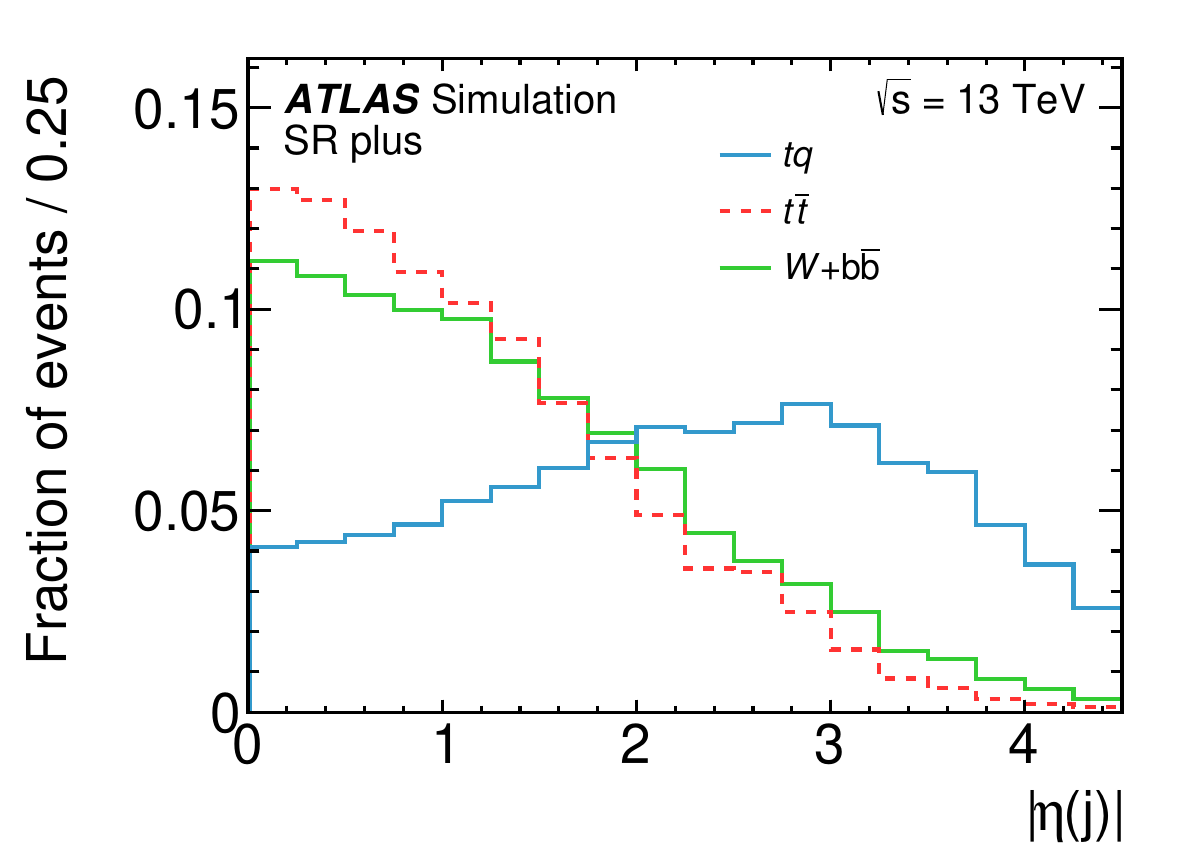}
\label{fig:input_var_shapes_etaj}
}
\caption{\label{fig:input_vars_shapes}
Probability densities of the two most discriminating input variables to the
NN in \srp.
\protect\subref{fig:input_var_shapes_mjb} The invariant mass $m(j b)$ of
the untagged jet and the $b$-tagged jet
and \protect\subref{fig:input_var_shapes_etaj}
the absolute value of the pseudorapidity of the untagged jet $|\eta(j)|$.
The distributions are shown for the $tq$ signal process, and the \(t\tbar\) and
the \wbb backgrounds.
Events beyond the $x$-axis range are included in the last bin.
}
\end{figure}

A single NN is trained using a sample of simulated events comprising both the positively and
negatively charged leptons, since the event kinematics of \tq and \tbarq production is very
similar.
This simple approach gives similar sensitivity as a scenario in which separate NNs are trained
in the \srp and \srm. The NN is trained against
all considered backgrounds with a fraction of 50\% signal events and 50\% background
events.
The different background processes are weighted relative to each other according to their
expected numbers of events.
NeuroBayes uses Bayesian regularisation techniques for the training process to improve
the generalisation performance and to avoid overtraining. The network infrastructure
consists of one input node for each input variable plus one bias node, followed by 22 nodes
arranged in a single hidden layer, and one output node that gives a
continuous output in the interval \((-1,+1)\).
As a non-linear activation function NeuroBayes uses the symmetric sigmoid function
\[ S(x) = \frac{2}{1+e^{-x}}-1\,, \]
which maps the interval $(-\infty, +\infty\,)$ to the interval $(-1,+1\,)$.
In the region close to zero, the sigmoid function has a linear response.
The final discriminant \dnn is obtained by linearly scaling the output of the NN to the interval $(0,1)$.

The probability densities of \dnn for the two SRs are shown in \Fig{\ref{fig:nn_templates}}
for the \tq signal process and the main backgrounds, namely the \ttbar and \wbb processes.
\begin{figure}[!t]
\centering
\subfloat[][]{
\includegraphics[width=0.47\textwidth]{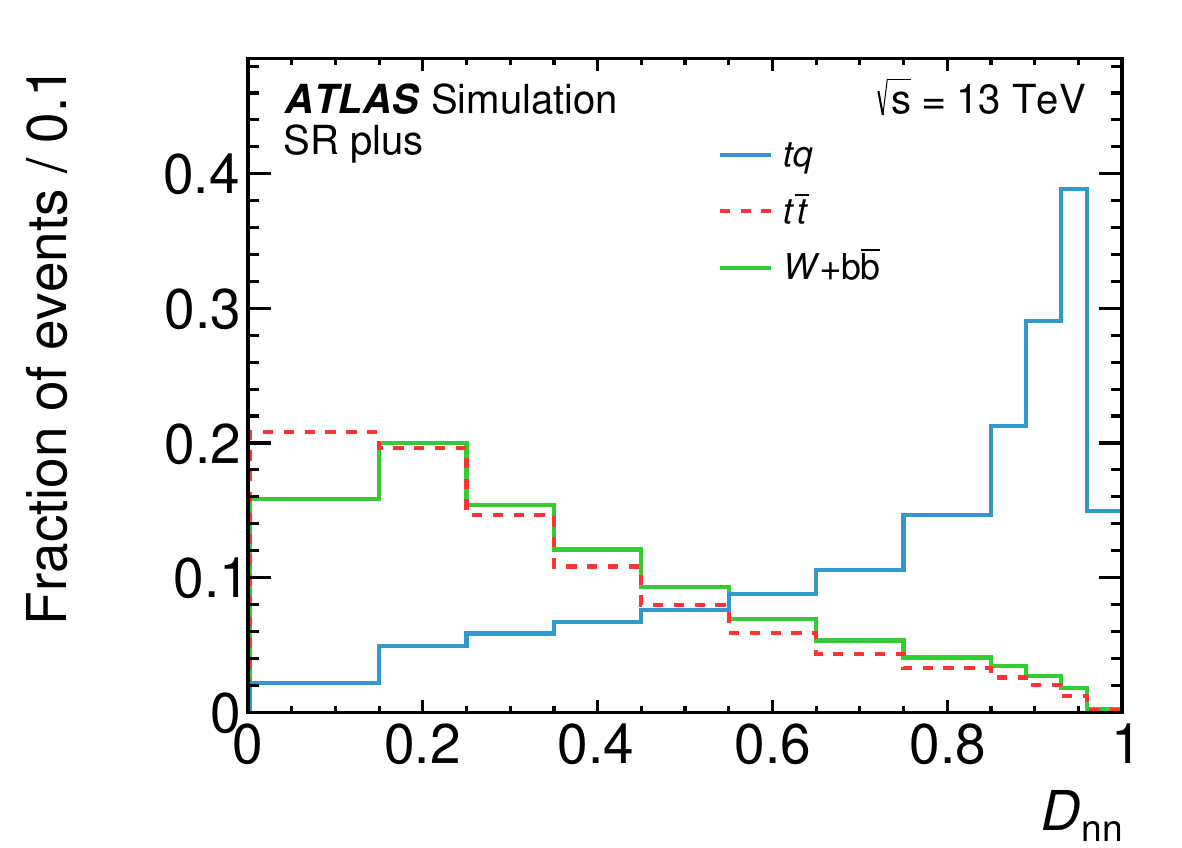}
\label{subfig:2j_plus_NNshape}
}
\quad
\subfloat[][]{
\includegraphics[width=0.47\textwidth]{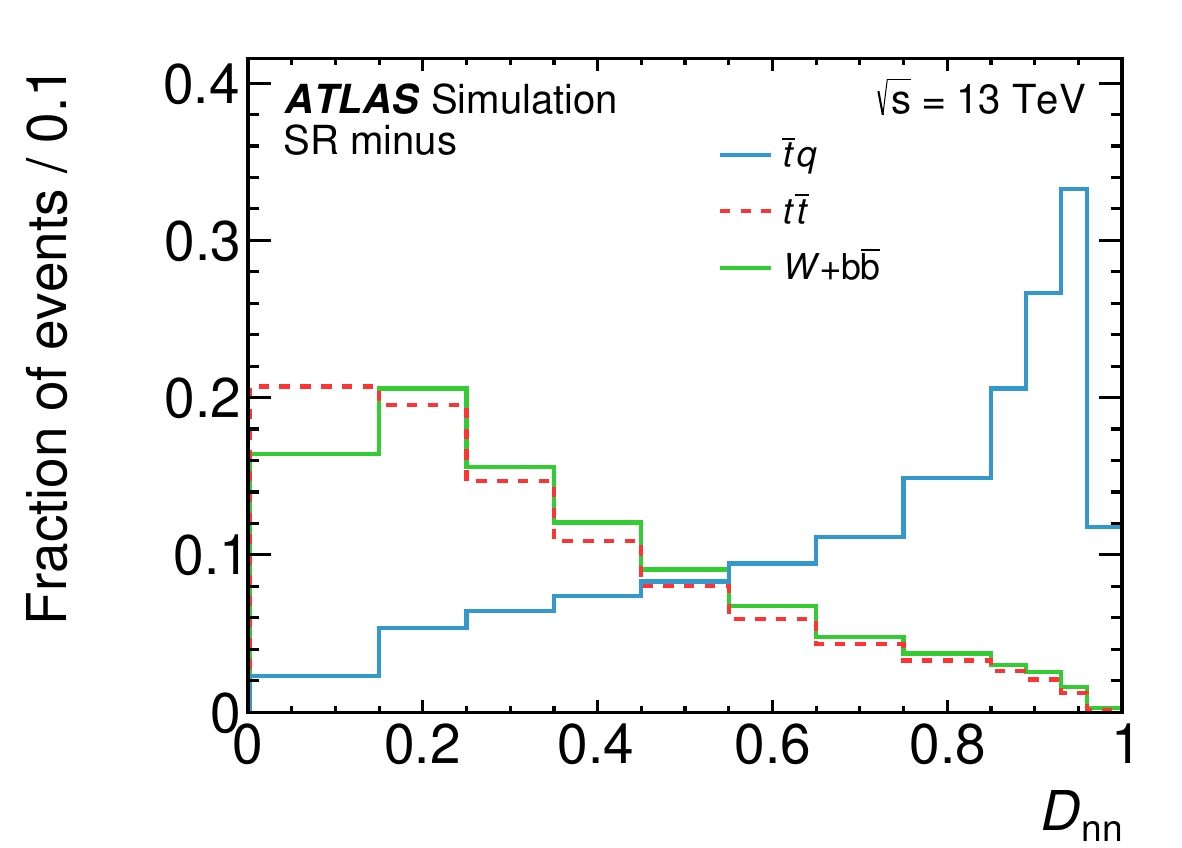}
\label{subfig:2j_minus_NNshape}
}
\caption{{\label{fig:nn_templates}
Probability densities of the NN discriminants for
the $tq$ and $\bar{t}q$ signal processes, and the \(t\tbar\) and \wbb backgrounds in
\protect\subref{subfig:2j_plus_NNshape} \srp and
\protect\subref{subfig:2j_minus_NNshape} \srm.
}
}
\end{figure}
Prior to the application of the NN to the observed collision data in the SRs,
the modelling of the input variables is checked. For this purpose, a preliminary estimate of the rate
of the multijet background is obtained by fitting the full $\met$ distribution for electron events
and fitting the \(\Delta\phi(\MET,\ell)\) distributions in \crmup and \crmum.
Since the resulting estimate of the multijet background is only a preliminary step towards
the final results, this fit is performed without using uncertainties other than the statistical data uncertainty
and the MC statistical uncertainties.
In the validation plots, the rates of all other processes including the signal process are set to their
predicted values.
The distributions of the eight most discriminating variables before performing the final maximum-likelihood fit
(pre-fit)
are shown in
\Figs{\ref{fig:input_vars_pre1}}{\ref{fig:input_vars_pre2}} for \srp.
In all cases, the model describes the observed distributions within the estimated uncertainties.
The pre-fit \dnn distributions are shown in \Fig{\ref{fig:nn_prefit}} for \srp and \srm.
\begin{figure}[!t]
\centering
\subfloat[][]{%
\includegraphics[width=0.46\textwidth]{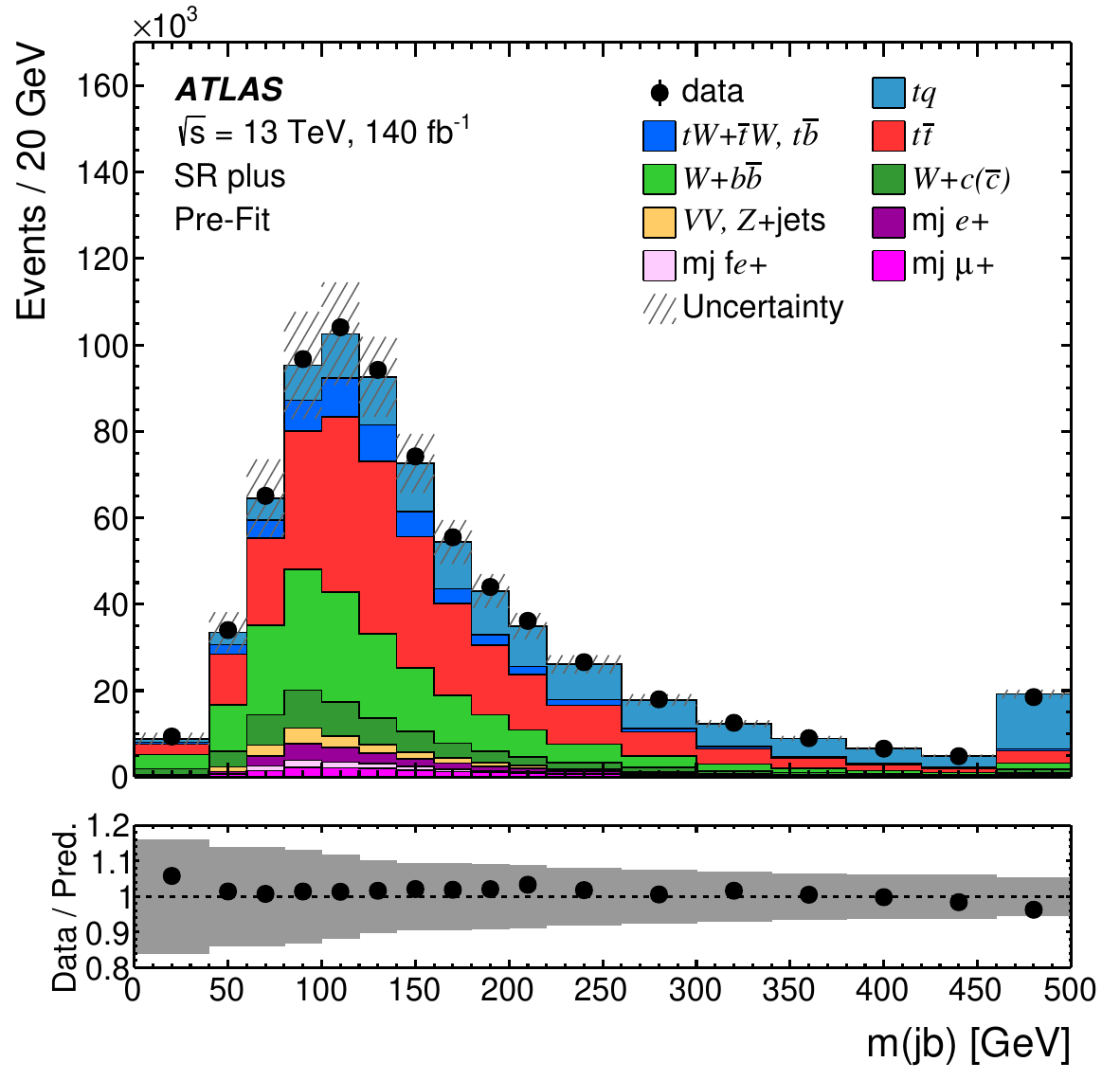}
\label{fig:input_vars_pre_mjb}
}
\,
\subfloat[][]{%
\includegraphics[width=0.46\textwidth]{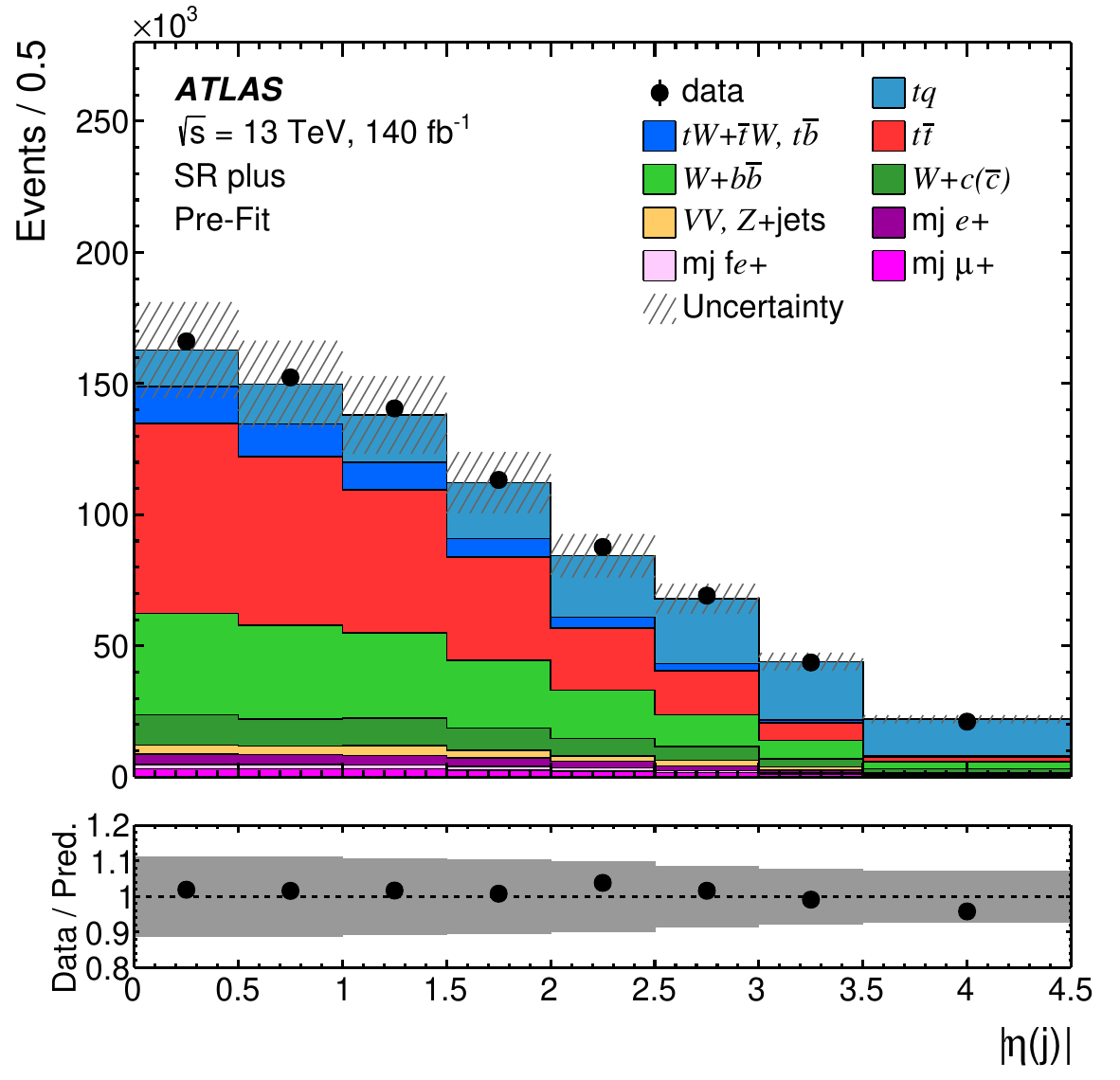}
\label{fig:input_vars_pre_etaj}
}\\
\subfloat[][]{%
\includegraphics[width=0.46\textwidth]{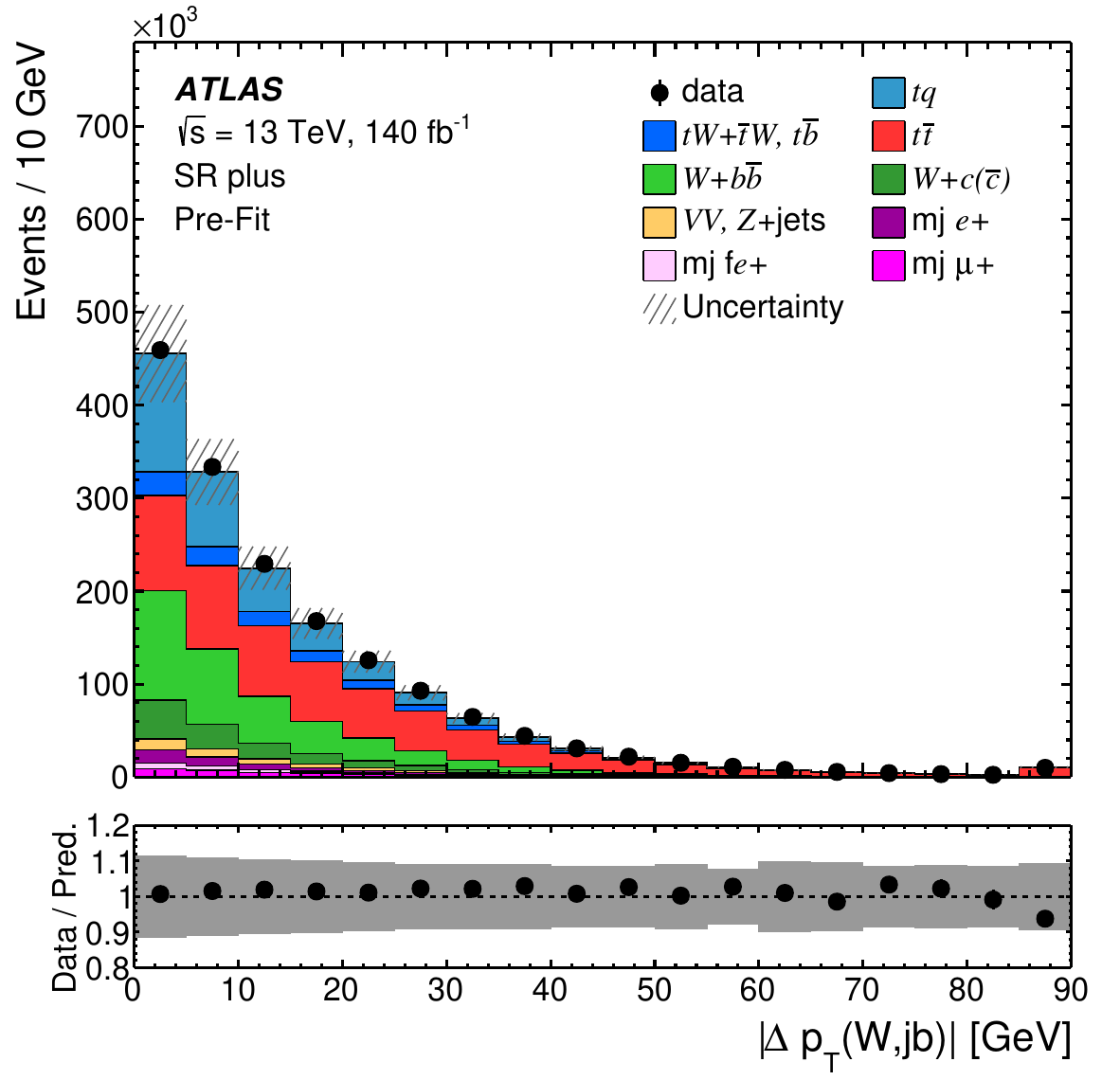}
\label{fig:input_vars_pre_dptWjb}
}
\,
\subfloat[][]{%
\includegraphics[width=0.46\textwidth]{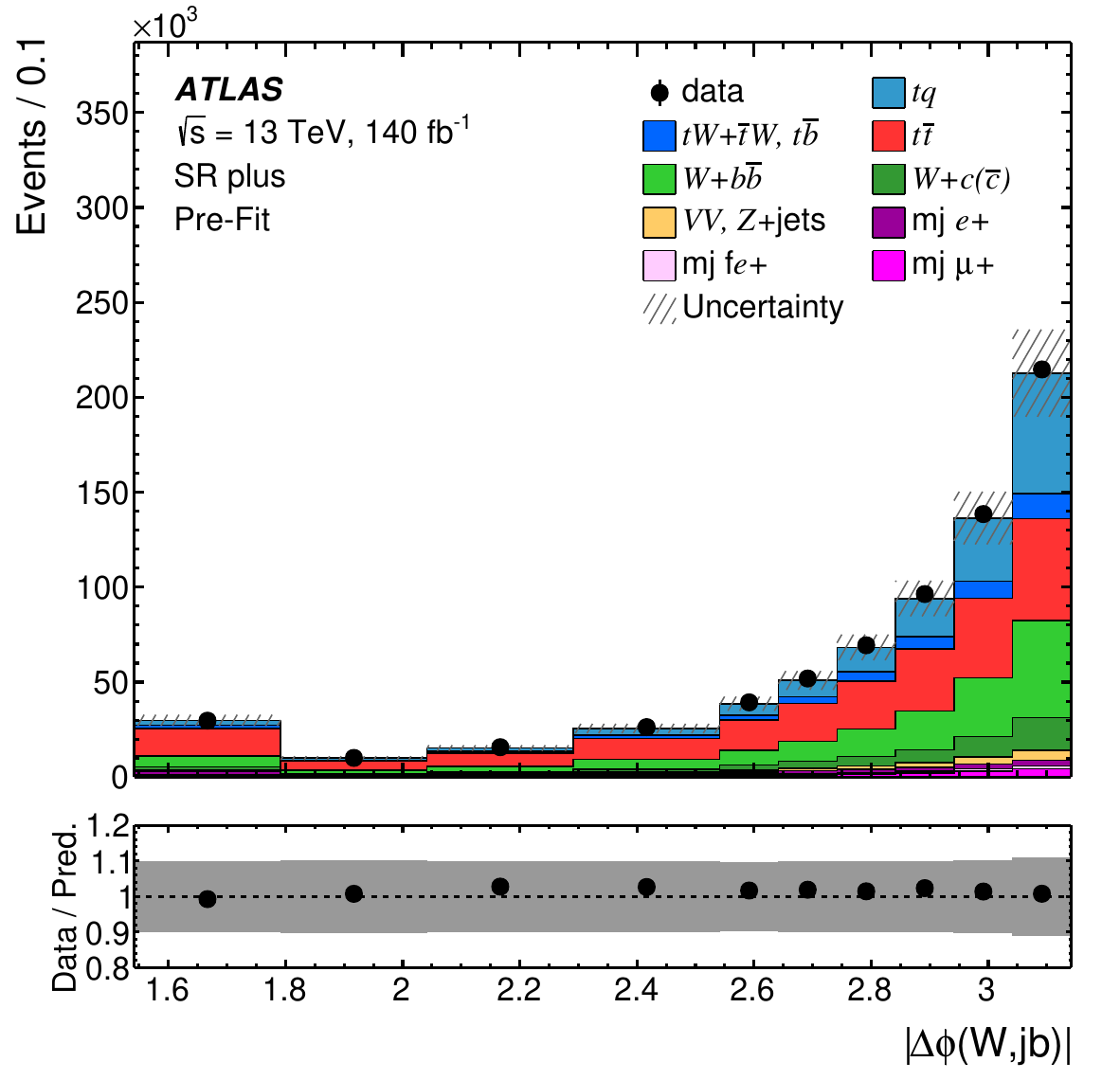}
\label{fig:input_vars_pre_dphiWjb}
}
\caption{\label{fig:input_vars_pre1}
Pre-fit distributions of the four most discriminating input variables to the
NN in \srp:
\protect\subref{fig:input_vars_pre_mjb}~the invariant mass~$m(j b)$ of
the untagged jet and the $b$-tagged jet,
\protect\subref{fig:input_vars_pre_etaj}~the absolute value of the pseudorapidity of the untagged jet~$|\eta(j)|$,
\protect\subref{fig:input_vars_pre_dptWjb}~the absolute value of the
difference in~$\pT$ between the reconstructed $W$~boson and the jet pair, and
\protect\subref{fig:input_vars_pre_dphiWjb}~the difference in azimuth angle between the reconstructed $W$~boson and the jet pair~\(|\Delta\phi(W, jb)|\).
The observed distributions (dots) are compared with the expected distributions (histograms) from simulated events. In these distributions, the signal contribution is shown stacked on top of contributions from all considered background processes.
All uncertainties considered in the analysis are included in the
hatched uncertainty band.
Events beyond the $x$-axis range are included in the last bin; the same applies to the first bin
of the \(|\Delta\phi(W, jb)|\) distribution in \protect\subref{fig:input_vars_pre_dphiWjb}.
The lower panel shows the ratio of data and the prediction; in this panel, the uncertainty is displayed as a grey band.
}
\end{figure}
\begin{figure}[p]
\centering
\subfloat[][]{%
\includegraphics[width=0.46\textwidth]{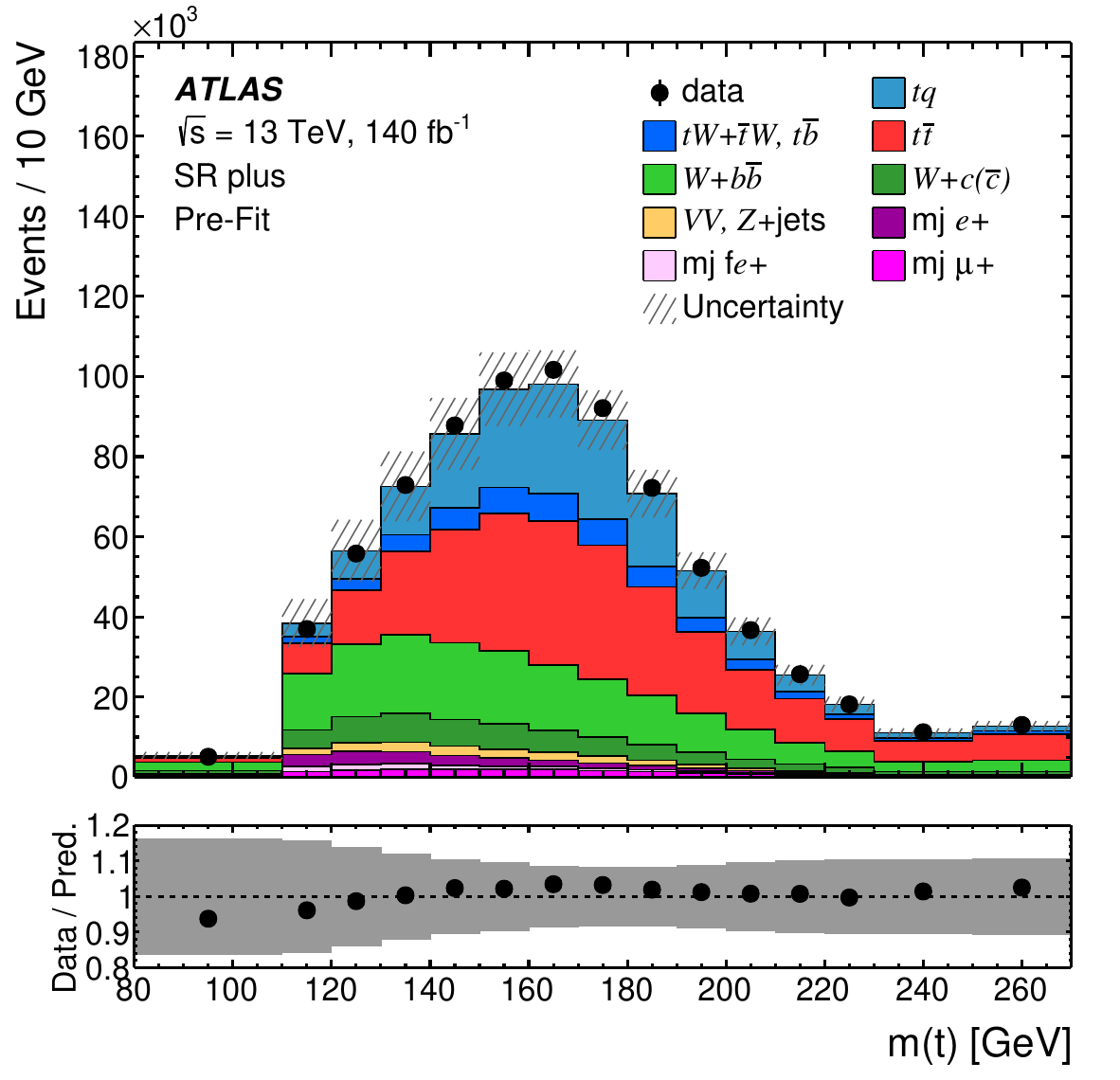}
\label{fig:input_vars_pre_mt}
}
\,
\subfloat[][]{%
\includegraphics[width=0.46\textwidth]{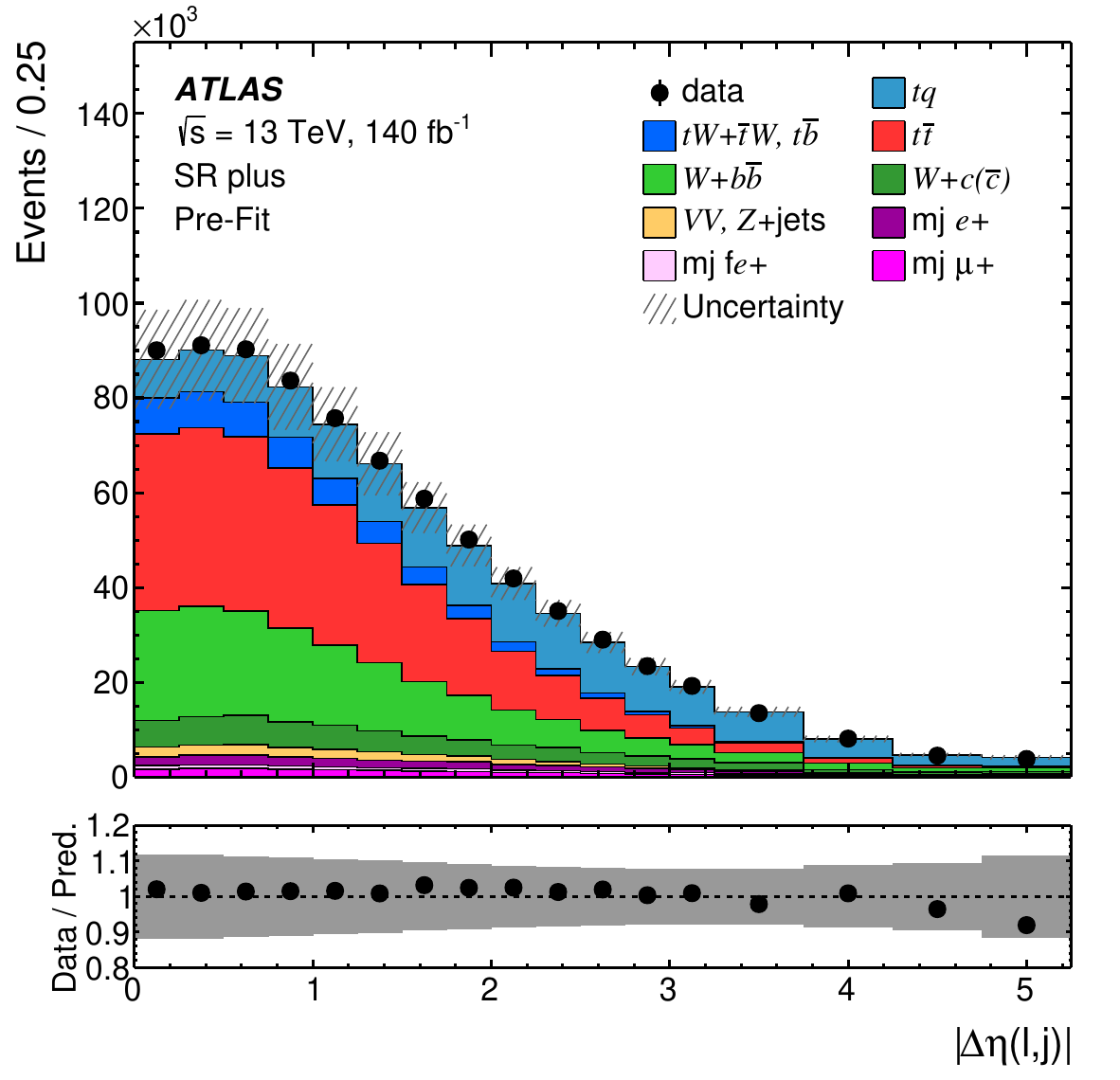}
\label{fig:input_vars_pre_detalepj}
}\\
\subfloat[][]{%
\includegraphics[width=0.46\textwidth]{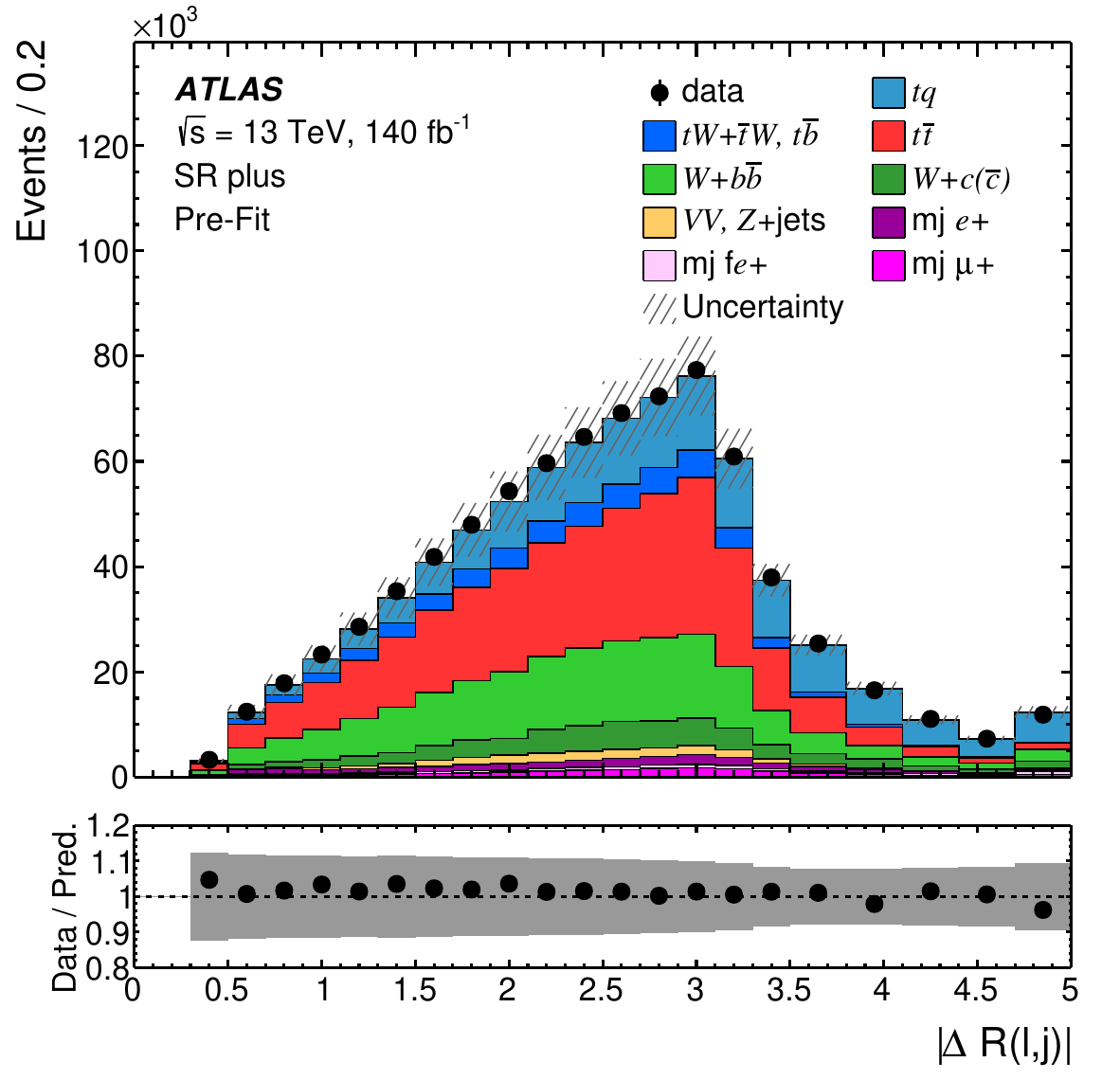}
\label{fig:input_vars_pre_drlepj}
}
\,
\subfloat[][]{%
\includegraphics[width=0.46\textwidth]{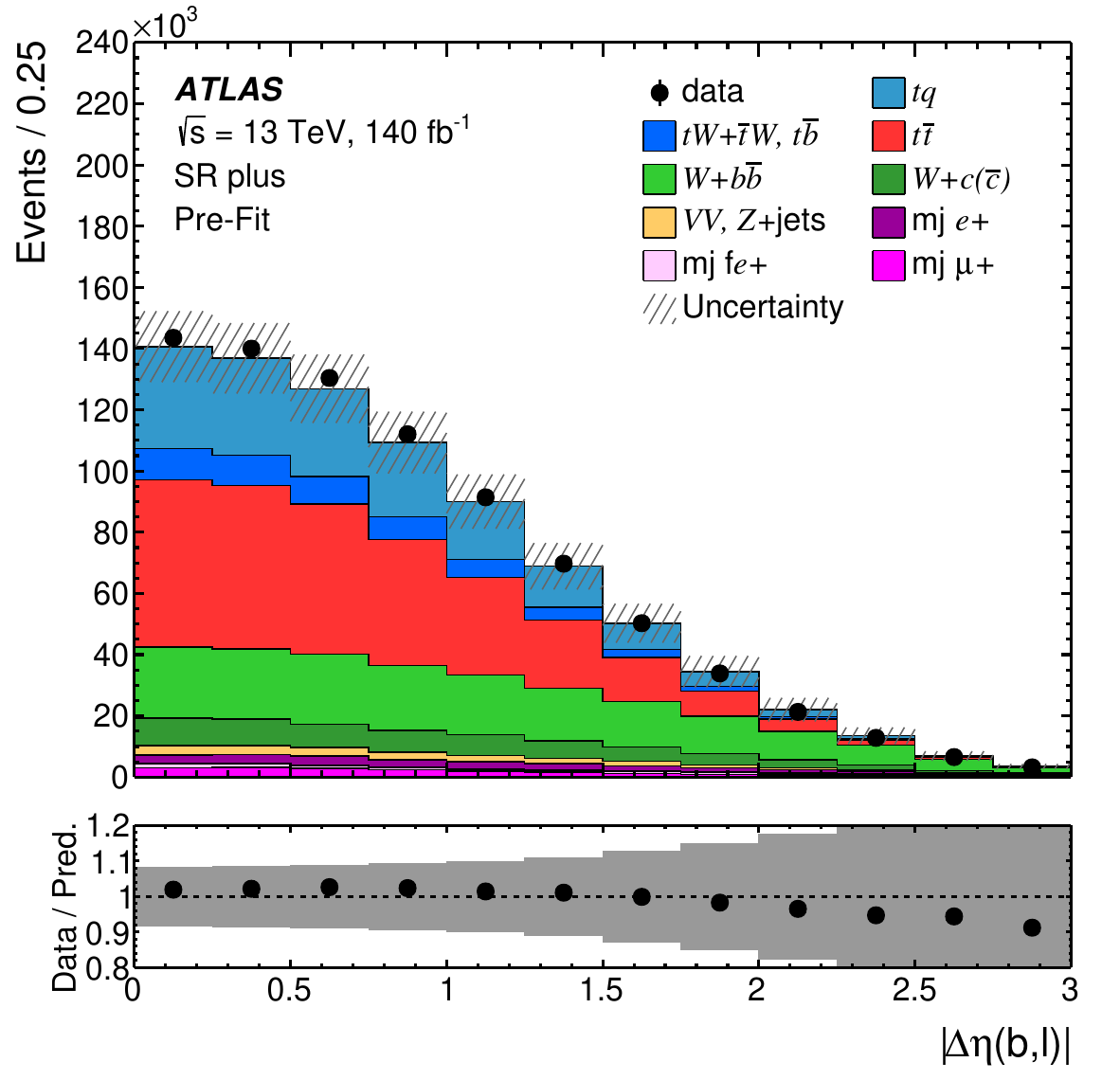}
\label{fig:input_vars_pre_detablep}
}
\caption{\label{fig:input_vars_pre2}
Pre-fit distributions of the four next most discriminating input variables to the NN in \srp:
\protect\subref{fig:input_vars_pre_mt}~the invariant mass of the reconstructed
top quark~\(m(t)\),
\protect\subref{fig:input_vars_pre_detalepj}~the absolute value of the difference in pseudorapidity between the charged lepton and
the untagged jet~\(|\Delta\eta(\ell ,j)|\),
\protect\subref{fig:input_vars_pre_drlepj}~the angular distance of the charged lepton and the untagged jet~\(\Delta R(\ell ,j)\), and
\protect\subref{fig:input_vars_pre_detablep}~the absolute value of the difference in pseudorapidity between the $b$-tagged jet
and the charged lepton~\(|\Delta\eta(b,\ell )|\).
The observed distributions (dots) are compared with the expected distributions (histograms) from simulated events.
In these distributions, the signal contribution is shown stacked on top of contributions from all considered background processes.
All uncertainties considered in the analysis are included in the
hatched uncertainty band.
Events beyond the \(x\)-axis range are included in the last bin. The lower panel shows the ratio of data and the prediction; in this panel, the uncertainty is displayed as a grey band.
}
\end{figure}
\begin{figure}[htbp]
\centering
\subfloat[][]{
\includegraphics[width=0.47\textwidth]{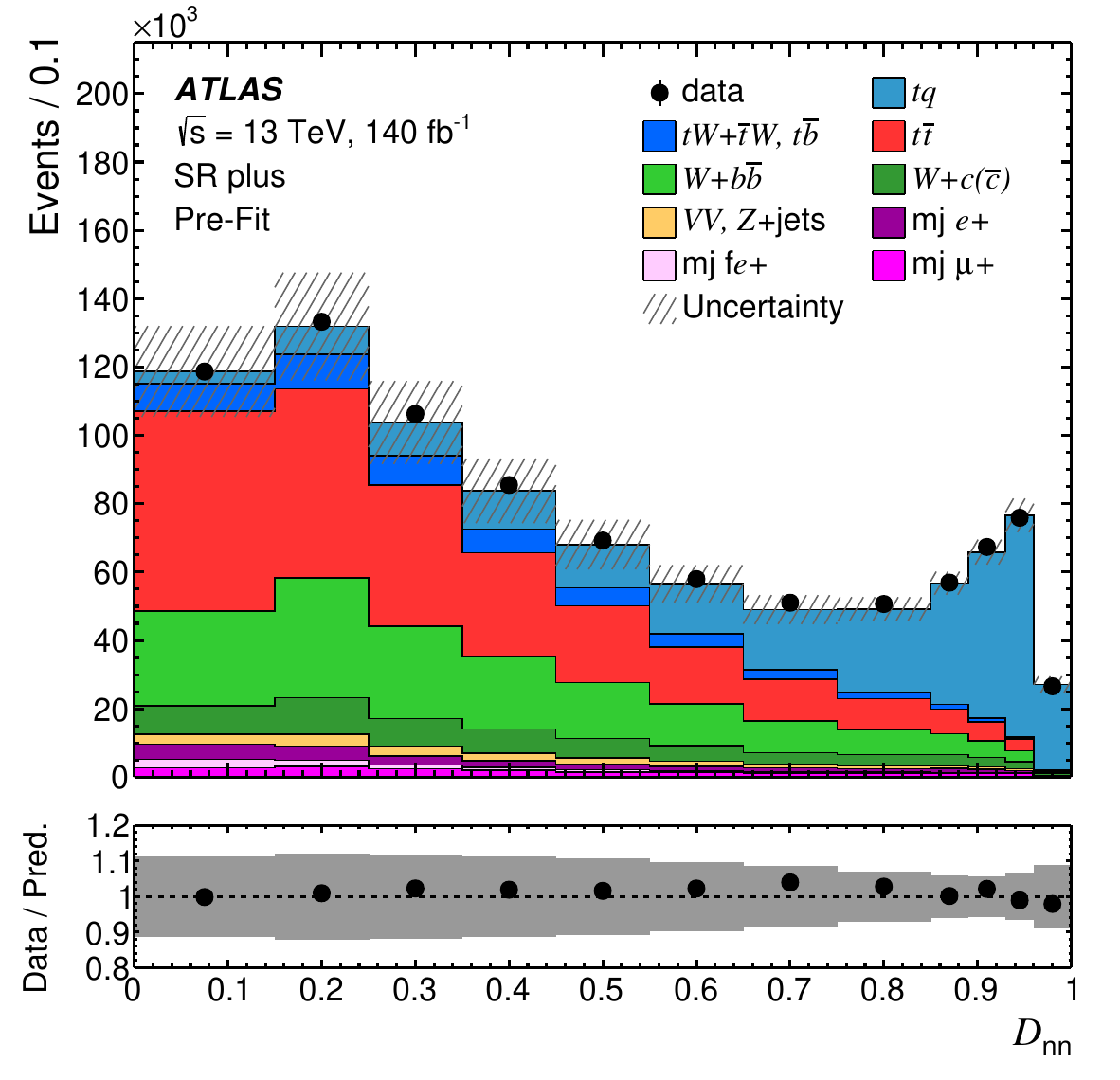}
\label{subfig:2j_plus_NNpre}
}
\quad
\subfloat[]{
\includegraphics[width=0.47\textwidth]{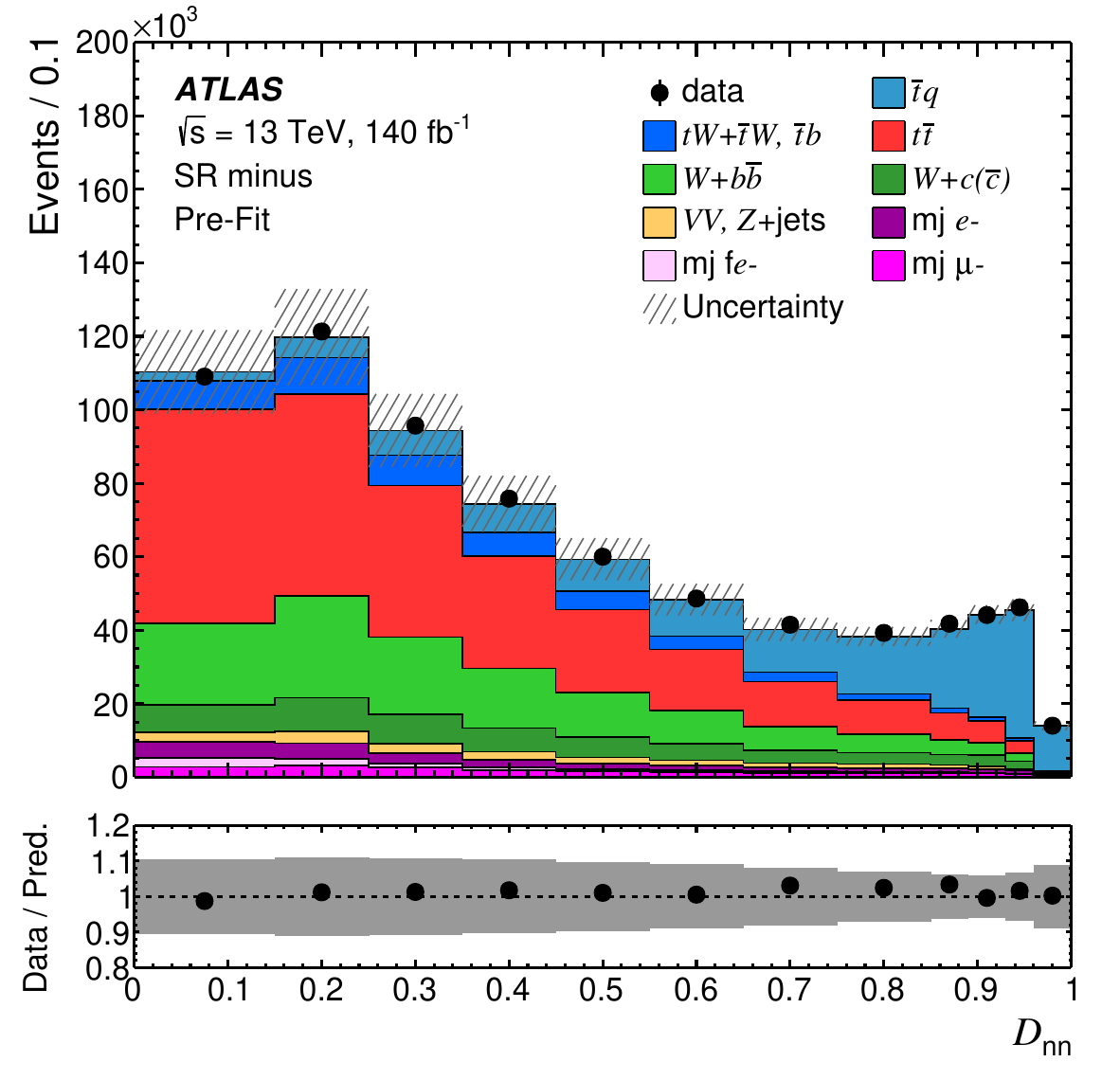}
\label{subfig:2j_minus_NNpre}
}
\caption{\label{fig:nn_prefit}
Pre-fit distributions of the \dnn in
\protect\subref{subfig:2j_plus_NNpre} \srp and
\protect\subref{subfig:2j_minus_NNpre} \srm. The observed distributions (dots) are compared with the expected distributions (histograms) from simulated events. In these distributions, the signal contribution is shown stacked on top of contributions from all considered background processes.
All uncertainties considered in the analysis are included in the
hatched uncertainty band. The lower panel shows the ratio of data and the prediction;
in this panel, the uncertainty is displayed as a grey band.
}
\end{figure}

%


\section{Systematic uncertainties}
%

%
%
Several sources of systematic uncertainty affect the expected event yield from signal and
background processes, and the shape of the NN discriminants used in the maximum-likelihood fits.
The systematic uncertainties are divided into two major categories.
Experimental uncertainties are associated with the reconstruction of the four-momenta of
final-state objects: electrons, muons, untagged jets, \btagged jets, and $\MET$.
The second category of uncertainties is related to the modelling of scattering processes.
All uncertainties are propagated through the analysis and their effects on the expected event
yields and discriminant distributions are accounted for
by including corresponding nuisance parameters in the fit.
In the following, the estimation of experimental and modelling uncertainties is explained in more detail.

\subsection{Experimental uncertainties}
The uncertainty in the integrated luminosity of the combined 2015--2018 data set is 0.83\% and
is based on a calibration of the luminosity scale using $x$--$y$ beam-separation
scans~\cite{DAPR-2021-01}. The luminosity uncertainty is applied to the expected signal and background
event yields except for the multijet background, which is estimated in a data-driven way.

Scale factors are applied to simulated events to correct for reconstruction, identification,
isolation and trigger performance differences between data and detector simulation for electrons and muons.
These scale factors and their systematic uncertainties, as well as the lepton momentum scale and resolution,
were assessed using $Z\rightarrow e^+e^-$ and $Z\rightarrow \mu^+\mu^-$ events in simulation
and data~\cite{EGAM-2018-01,MUON-2018-03,EGAM-2021-02,MUON-2022-01}.
The probability of charge misidentification of reconstructed electrons is estimated from simulated events
to be \(1.9\times10^{-3}\). The net effect is a migration of events at the
level of \(6\times10^{-4}\) from the \srp to the \srm (see \Tab{\ref{tab:postfityields}}).
An uncertainty of 100\% on this rate is applied in the maximum-likelihood fits and is taken to be anticorrelated
between the \tq and the \tbarq processes.

The jet energy scale (JES) was calibrated using a combination of test-beam data, simulation and
in situ techniques~\cite{JETM-2018-05}.
The JES is parameterised in bins of jet \pt and $\eta$.
Its uncertainty is decomposed into a set of 30 uncorrelated components,
of which 29 are non-zero in a given event depending on the type of simulation used.
Sources of uncertainty contributing to the JES uncertainties computed using the detector position $\eta_\mathrm{det}$ include the $\eta$ intercalibration of forward jets within $0.8< |\eta_\mathrm{det}| < 4.5$ with those in the
central barrel region ($|\eta_\mathrm{det}| < 0.8$), pile-up modelling, jet flavour composition and response, differences between jets induced by \bquarks and those from gluons or light-quarks,
single-particle response, detector modelling, non-closure, and effects of jets not fully contained
within the calorimeter.

The uncertainty of the jet energy resolution (JER) is evaluated by smearing jet energies according to a Gaussian
function with width \(\sigma_\mathrm{smear}\)~\cite{JETM-2018-05}.
Thirteen orthogonal components account
for jet $\pt$- and $\eta$-dependent differences between simulation and data
that were determined using dijet events and noise measurements based on random cones.
The smearing is applied to simulated events, used to build the fit model, if the resolution in data is larger than in MC simulation,
and to pseudo-data, obtained from simulated events, when the resolution is larger in simulation than in collision data.
The JER uncertainties are defined by combining both variations and thereby taking
the anticorrelation between different components into account.
The nominal data remain unchanged.
The uncertainty in the efficiency to satisfy the JVT requirement for pile-up suppression
was derived in \(Z(\rightarrow \mu^+\mu^-)\)+jets events and is also considered~\cite{PERF-2014-03}.
The uncertainty in \met due to a possible miscalibration of its soft-track component was derived
from data--simulation comparisons of the \pT balance between the hard and soft \met
components~\cite{PERF-2016-07}.

The \btag requirement made in the measurement requires the consideration of uncertainties in the \btag efficiency of
true \bjets and in the mistagging rates of light-quark jets and \(c\)-jets.
The \btag efficiency is measured in dileptonic \(t\tbar\) events. Differences between data
and detector simulation are corrected by \pt-dependent scale factors applied to simulated events.
The uncertainty in the scale factors is decomposed into 45 orthogonal components~\cite{FTAG-2018-01}.
The uncertainties depend on the \pt of the \bjets and are propagated through the analysis as weights.
The rate of mistagging \(c\)-jets as \bjets was measured in semileptonic \(t\tbar\) events, where one
of the $W$ bosons decays into an electron or a muon and a neutrino and the other decays into a
quark--antiquark pair~\cite{FTAG-2020-08}.
This event sample allows the measurement to utilise the relatively large and known
\(W\rightarrow cs\) branching ratio. The mistagging rate of \(c\)-jets depends on the jet \pt and
has a total uncertainty in the range of 3\%-17\%. The uncertainties are decomposed into 20
orthogonal components.
The misidentification rate of light-quark jets was evaluated based on the techniques described in
Ref.~\cite{ATLAS:2023lwk}.
The resulting calibration factors are in the range of about
1.5 to 3 with uncertainties up to 50\%. The uncertainties are decomposed into 20 independent
eigenvectors.

\subsection{Modelling uncertainties}
Uncertainties in the theoretical cross-sections are evaluated for the top-quark background processes
($t\tbar$, $tW$ and $t\bbar$) as quoted in Section~\ref{sec:samples_top_quark}.
Due to the tight \btag requirement the largest contribution to the \Wjets background comes
from \wbb production in which the \bquarks are produced via gluon splitting
(\(g\rightarrow b\bbar\)). An uncertainty of \(\pm 40\%\) is assigned to the expected rate of
this process, covering differences seen in previous measurements~\cite{TOPQ-2018-06} between the \SHERPA prediction and
ATLAS collision data.
The contribution of the associated production of a \(W\) boson and light-quark jets to the
expected event yield is much smaller and therefore this contribution is merged with
the \wbb process in the statistical analysis. The same uncertainty of \(\pm 40\%\)
is assigned to it.
Events in which a \(W\) boson is produced in association with \(c\)-jets are mainly due to
the $sg\rightarrow W^{-}c$ and $\bar{s}g\rightarrow W^{+}\bar{c}$ scattering processes.
An uncertainty of \(\pm 20\%\) is assigned to the rate of \(W\)+\(c\)-jets production.
The same uncertainty is applied to the rate of the combined process of \Zjets and diboson production.
The fit result and its uncertainty depend only marginally on the specific assignment of the uncertainties
(\(\pm40\%\) and \(\pm20\%\), respectively) in the \wbb, \Zjets, and diboson cross-sections.
In the maximum-likelihood fit, separate nuisance parameters are used for the cross-section uncertainties
of \wbb production and \(W\)+\(c\)-jets production in the regions with
positive and negative charge.

Uncertainties in modelling parton showers and hadronisation are assigned to the $tq$ signal
and the top-quark background processes (\ttbar, $tW$ and $t\bar{b}$ production) by comparing
the nominal samples with alternative samples for which \POWHEGBOX[v2] was interfaced to
\HERWIG[7.2.1]~\cite{Bahr:2008pv,Bellm:2015jjp} (for \ttbar production) or \HERWIG[7.1.6] (for $tq$, $tW$ and $t\bar{b}$ production)
instead of \PYTHIA[8.230].
The uncertainties are considered to be uncorrelated for the different scattering processes,
namely the $tq$ signal process and the three top-quark background processes.
In the statistical analysis, normalisation and shape effects are decorrelated as well.

Uncertainties related to the choice of \(\muR\) and \(\muF\) for the matrix-element
calculations are evaluated by varying the scales independently by factors of
\(2\) and \(0.5\), separately for each of the top-quark production processes and
for \Wjets production. The scale variations are implemented as generator weights
in the nominal sample. These weights are propagated through the entire analysis.

The uncertainty in matching the NLO matrix elements to the parton shower when generating
\(t\tbar\) and \tq events is evaluated by comparing the nominal samples of simulated events
to samples with an alternative setting of the \(\pt^{\mathrm{hard}}\) parameter in the
matching code. This parameter regulates the
definition of the vetoed region of the parton shower and is thus important to avoid overlap in the
phase space filled by \POWHEG and \PYTHIA.
The nominal setting of \(\pt^{\mathrm{hard}}=0\) imposes a veto based on the \pT of the gluon
emission produced by \POWHEG, while the alternative setting \(\pt^{\mathrm{hard}}=1\) leads to
a veto defined by the minimal \pT among all final-state partons.
This estimate of the uncertainty follows the description in Ref.~\cite{Hoche:2021mkv}
and was studied by the ATLAS Collaboration in \Refn{\cite{ATL-PHYS-PUB-2023-029}}.
The uncertainty in the choice of the $h_\mathrm{damp}$ parameter for the \(t\tbar\) event
generation is estimated by using an additional \ttbar sample produced with the $h_\mathrm{damp}$
parameter set to $3\times m_t$,
while keeping all other generator settings the same as for the nominal sample of events.

Uncertainties in the amount of initial-state and final-state radiation are assessed for the
top-quark production processes by varying the parameter \texttt{Var3c} of the A14 parton-shower tune
within the uncertainties of the tune and, for final-state radiation, by varying the
renormalisation scale \(\muR\), at which the strong coupling constant $\alpha_\mathrm{s}$ is evaluated,
by factors of $0.5$ and $2.0$. The two variations are handled independently.
The uncertainty due to the scheme for removing the overlap of the \(tW\) process with
\(\ttbar\) production is evaluated by comparing the nominal sample, using the diagram-removal scheme,
with a sample produced with an alternative scheme (diagram subtraction)~\cite{Frixione:2008yi}.

In all uncertainty evaluations mentioned above the alternative samples or reweighted samples are
normalised to the total cross-section of the nominal samples.

Uncertainties in the PDFs are evaluated for the top-quark production processes using the
PDF4LHC15 prescription with 30 eigenvectors~\cite{Butterworth:2015oua}.
Simulated events are reweighted to the central value and the eigenvectors of the combined PDF set.
Systematically varied templates are constructed by taking the differences between the samples
reweighted to the central value and those reweighted to the eigenvectors.
In the likelihood fit, the PDF uncertainties are treated as correlated across the
top-quark production processes.

The uncertainty in the multijet background is evaluated by modifying the selection
criteria for jet-electron and non-prompt-muon candidates. Two alternative
selections of jet-electron candidates are defined by varying the requirement on the energy fraction measured in
the electromagnetic calorimeter, leading to two alternative shapes of the \dnn distributions
for the multijet background in the SRs. In the statistical analysis, these shapes are used as
``up'' and ``down'' variations of a single nuisance parameter.
For non-prompt-muon candidates a single variation is defined by varying the isolation criteria.

To account for differences in the pile-up distribution between simulation and data, the pile-up profile
in the simulation is corrected to match the one in data.
The uncertainty associated with the correction factor is applied in the measurement as a variation
of the event weight.

The uncertainties due to the finite number of simulated events, also called the MC statistical uncertainty,
is accounted for by adding a nuisance parameter for each bin of the \dnn distributions and
the distributions in the CRs, implementing the Barlow--Beeston approach~\cite{Barlow:1993dm}.


\section{Measurement results}
\label{sec:results}

%
The cross-sections \sigt and \sigat are determined in a simultaneous binned profile maximum-likelihood fit.
To properly account for the correlations of systematic uncertainties
when forming the sum and the ratio of \sigt and \sigat, $\sigma(\tq+\tbarq)$ and
\rt are measured in a second fit in which the parameterisation of the signal strength
parameters is modified accordingly, while all other parameters of the fit setup are kept the same.
The fitted distributions are the \dnn distributions in \srp and \srm,
the \(\Delta\phi(\MET,\ell)\) distributions in the \crmup and the \crmum,
and the event yields in the
CR \texttt{B-e-plus}, the CR \texttt{B-e-minus}, the CR \texttt{EC-e-plus} and
the CR \texttt{EC-e-minus}.
In both fits, the event yields of the multijet background are left floating,
while the yields of all other backgrounds are constrained to their predictions within the
associated uncertainties.

The likelihood function is constructed as a product of Poisson probability
terms over all considered bins.
The fitted event yields in the bins depend on nuisance parameters that include the
effects of systematic uncertainties.
Each nuisance parameter, except those representing the MC statistical uncertainties,
is constrained by a Gaussian term in the likelihood function.
Some systematically varied discriminant distributions are smoothed and nuisance parameters
of systematic uncertainties with negligible impact are entirely removed to
reduce spurious effects in the minimisation, improve the convergence of the fit,
and reduce the computing time.
Normalisation and shape effects of a source of systematic uncertainty are treated
separately in this removal process.
Single-sided systematic variations are turned into symmetric variations by taking the
full difference in event yield and shape between the nominal model and the alternative
model and mirroring this difference in the opposite direction.
For most sources with two variations, their effects are made symmetric by using the average
deviation from the nominal prediction.
Exceptions are the uncertainties in the JER and in the jet-electron model,
for which the asymmetric variations are kept because the underlying effects are known to be
asymmetric.
No significant pulls of nuisance parameters are observed.

The total cross-sections for \tq and \tbarq production are measured to be
\[
\sigt  = 137^{+8}_{-8}\,\mathrm{pb} \ \ \ \ \mathrm{and} \ \ \ \
\sigat =  84^{+6}_{-5}\,\mathrm{pb}.
\]
The NNLO predictions for these cross-sections (see \Sect{\ref{sec:samples_top_quark}})
agree very well with the measurements.
The relative precision reached is \(+5.9\%\) and \(-5.5\%\) for \(\sigt\) and
\(+6.6\%\) and \(-6.2\%\) for \(\sigat\).
The fits to the observed data for the combined \tq and \tbarq cross-section and \rt give
the following results:
\[
\sigtat = 221^{+13}_{-13}\,\mathrm{pb} \ \ \ \ \mathrm{and} \ \ \ \
\rt = 1.636^{+0.036}_{-0.034}
\]
with a relative precision of \(+6.1\%\) and \(-5.7\%\) for the combined cross-section and
\(+2.2\%\) and \(-2.1\%\) for \rt.
The global goodness of fit is evaluated with the saturated model~\cite{Workman:2022ynf} yielding a
\(p\)-value of \SI{76}{\percent}.
Table~\ref{tab:grouped_impact} provides a breakdown of the uncertainties categorised in groups according to
different sources.
The impact of a particular group of uncertainties is evaluated by performing an alternative likelihood fit in which
the nuisance parameters related to the sources of uncertainty under investigation are fixed to their
best-fit values as obtained from the nominal fit. The squared impact of the considered group of uncertainties is
determined as the difference between the square of the nominal total uncertainty and the square of the uncertainty
obtained from the alternative fit.
For the measured cross-sections, the uncertainties in the signal modelling are the dominating ones,
while they largely cancel out for the measurement of \rt.
The data statistical uncertainty is very small compared with the systematic uncertainties.
Since many uncertainties largely cancel out when forming the ratio, the uncertainty in \rt
is much reduced compared with the uncertainties in the cross-sections.

The eight single most important systematic uncertainties in the cross-section measurements
are listed in Table~\ref{tab:single_impact_xs}.
\begin{table}[!tb]
\centering
\begin{tabular}{lcccc}
\toprule
Uncertainty group & \(\Delta\sigt/\sigt\) & \(\Delta\sigat/\sigat\) & \(\Delta\sigtat/\sigtat\) & \(\Delta\rt/\rt\) \\
\midrule
Data statistics      &\(+0.4\) / \(-0.4\)& \(+0.5\) / \(-0.5\)&    \(+0.3\) / \(-0.3\) &    \(+0.6\) / \(-0.6\)\\
Signal modelling     &\(+4.9\) / \(-4.5\)& \(+5.2\) / \(-4.8\)&    \(+5.0\) / \(-4.6\) &    \(+0.9\) / \(-0.9\)\\
Background modelling &\(+1.8\) / \(-1.6\)& \(+2.1\) / \(-1.9\)&    \(+1.8\) / \(-1.6\) &    \(+1.5\) / \(-1.4\)\\
MC statistics        &\(+1.0\) / \(-1.0\)& \(+1.4\) / \(-1.3\)&    \(+1.1\) / \(-1.0\) &    \(+0.8\) / \(-0.8\)\\
PDFs                 &\(+0.4\) / \(-0.4\)& \(+1.2\) / \(-1.0\)&    \(+0.6\) / \(-0.6\) &    \(+0.9\) / \(-0.8\)\\
Jets                 &\(+2.2\) / \(-2.0\)& \(+3.0\) / \(-2.7\)&    \(+2.5\) / \(-2.2\) &    \(+1.0\) / \(-0.9\)\\
\btag                &\(+1.6\) / \(-1.5\)& \(+1.7\) / \(-1.5\)&    \(+1.6\) / \(-1.5\) &    \(+0.2\) / \(-0.1\)\\
Leptons              &\(+1.1\) / \(-1.0\)& \(+1.1\) / \(-1.0\)&    \(+1.1\) / \(-1.0\) &    \(+0.1\) / \(-0.1\)\\
Luminosity           &\(+0.9\) / \(-0.8\)& \(+0.9\) / \(-0.9\)&    \(+0.9\) / \(-0.8\) &    \(<0.1\)\\
\midrule
Total                &\(+5.9\) / \(-5.5\) & \(+6.6\) / \(-6.2\) &  \(+6.1\) / \(-5.7\) &    \(+2.2\) / \(-2.1\) \\
\bottomrule
\end{tabular}
\vspace*{2mm}
\caption{\label{tab:grouped_impact}The impact of different groups of systematic uncertainties
on \sigt, \sigat, \sigtat and \rt given in~\%.}
\end{table}
\begin{table}[!tb]
\centering
\begin{tabular}{lccc}
\toprule
Systematic uncertainty                   & \(\Delta\sigt/\sigt\) & \(\Delta\sigat/\sigat\) &
\(\Delta\sigtat/\sigtat\) \\
\midrule
\tq matching scale definition, rate       & \(+3.1\) / \(-2.9\) & \(+2.8\) / \(-2.6\) & \(+2.9\) / \(-2.8\) \\
\tq parton shower, rate                   & \(+2.6\) / \(-2.5\) & \(+3.3\) / \(-3.2\) & \(+2.9\) / \(-2.8\) \\
\tq final-state radiation                 & \(+2.1\) / \(-2.0\) & \(+2.2\) / \(-2.1\) & \(+2.1\) / \(-2.0\) \\
\tq matching scale definition, shape      & \(+1.6\) / \(-1.5\) & \(+1.2\) / \(-1.2\) & \(+1.5\) / \(-1.4\) \\
JES \(\eta\) intercalibration modelling   & \(+1.2\) / \(-1.2\) & \(+1.6\) / \(-1.5\) & \(+1.4\) / \(-1.3\) \\
\btag NP B1                               & \(+1.0\) / \(-0.9\) & \(+1.0\) / \(-1.0\) & \(+1.0\) / \(-0.9\) \\
\btag NP B0                               & \(+1.0\) / \(-0.9\) & \(+1.0\) / \(-1.0\) & \(+1.0\) / \(-0.9\) \\
Luminosity                                & \(+0.9\) / \(-0.8\) & \(+0.9\) / \(-0.9\) & \(+0.9\) / \(-0.8\) \\
\bottomrule
\end{tabular}
\vspace*{2mm}
\caption{\label{tab:single_impact_xs}The impact of the eight most important systematic uncertainties
on \sigt, \sigat and \sigtat given in \%. The sequence of the uncertainties
is given by the impact on \(\sigtat\).}
\end{table}
The four most important systematic uncertainties are due to the modelling of the \tq process
with an event generator.
The single largest uncertainty is the rate effect due to the definition of the matching scale of the
\tq process.
The eight most important systematic uncertainties in the \rt measurement are listed in
Table~\ref{tab:impact_rt}.
\begin{table}[!tb]
\centering
\begin{tabular}{lc}
\toprule
Systematic uncertainty                      & \(\Delta\rt/\rt\) \\
\midrule
\(W^-+c(\bar{c})\) cross-section             & \(+0.8\) / \(-0.8\) \\
\tq parton shower, rate                      & \(+0.7\) / \(-0.7\) \\
\(W^++c(\bar{c})\) cross-section	     & \(+0.5\) / \(-0.5\) \\
PDF eigenvector 09                           & \(+0.5\) / \(-0.5\) \\
MC statistical uncertainty in \dnn bin 10 of \srm        & \(+0.4\) / \(-0.4\) \\
JES \(\eta\) intercalibration modelling      & \(+0.4\) / \(-0.4\) \\
\tq matching scale definition, shape         & \(+0.4\) / \(-0.4\) \\
PDF eigenvector 05                           & \(+0.4\) / \(-0.4\) \\
\bottomrule
\end{tabular}
\vspace*{2mm}
\caption{\label{tab:impact_rt}The impact of the eight most important systematic uncertainties
on \rt in \%.}
\end{table}
As a cross-check, the selected sample of events was split according to the lepton flavour
into an electron and a muon sample and the measurements were repeated,
leading to results compatible with the nominal analysis.

The \dnn distributions after performing the fit are shown for both of the SRs in
\Fig{\ref{fig:nn_postfit}}.
\begin{figure}[!tbp]
\centering
\subfloat[][]{
\includegraphics[width=0.47\textwidth]{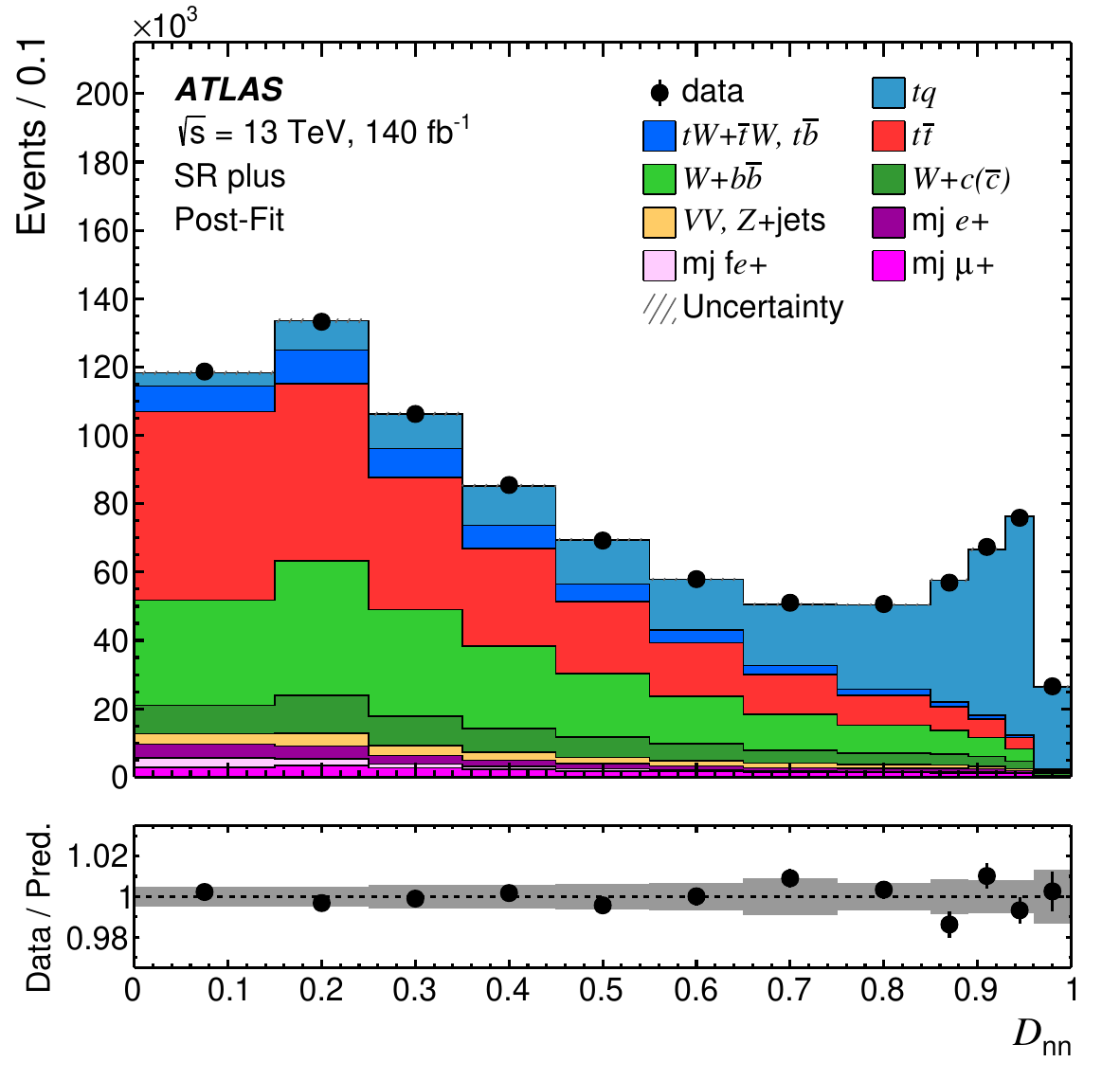}
\label{subfig:2j_plus_NNpost}
}
\quad
\subfloat[]{
\includegraphics[width=0.47\textwidth]{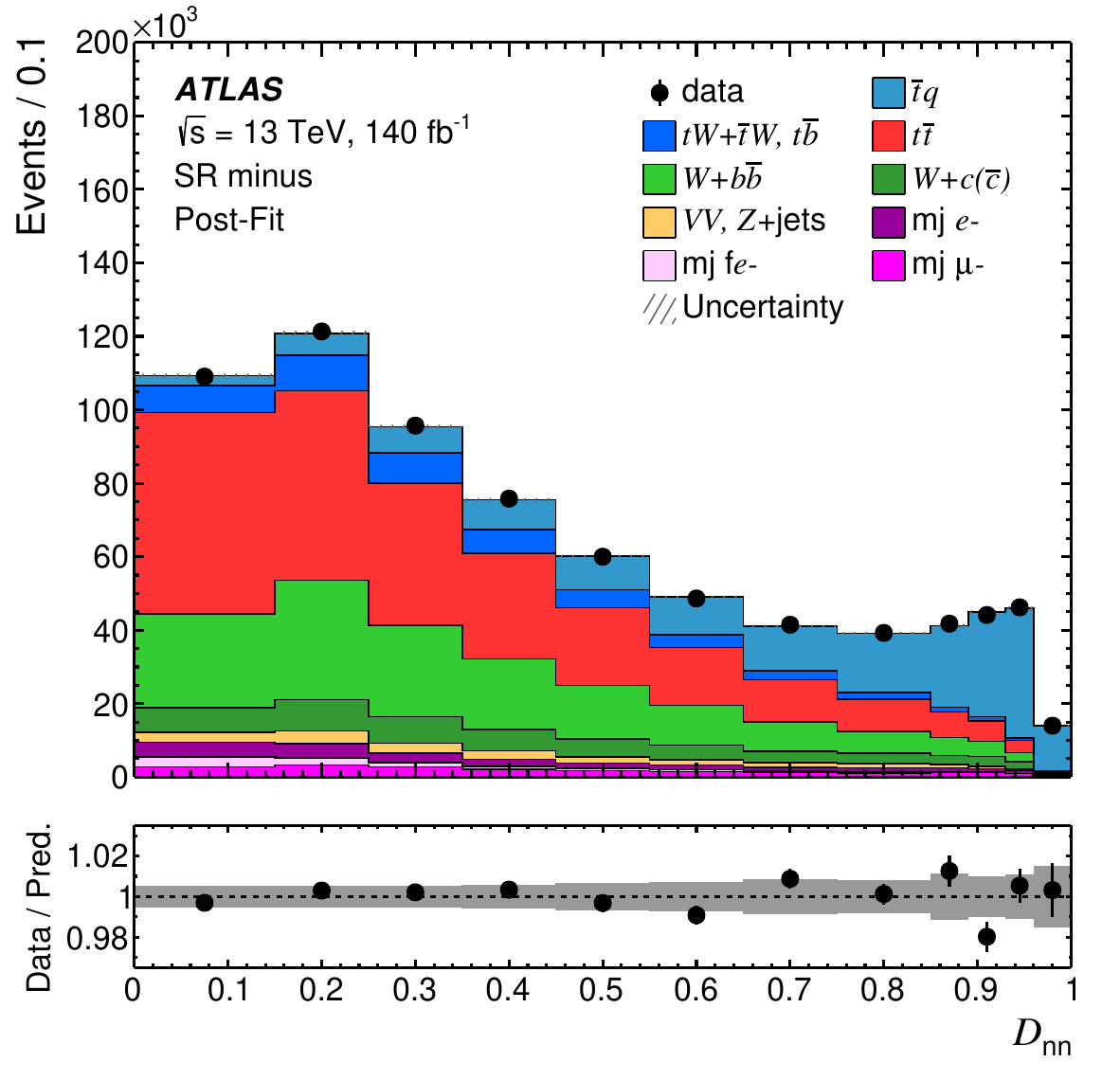}
\label{subfig:2j_minus_NNpost}
}
\caption{\label{fig:nn_postfit}
The observed \dnn distributions (dots) for
\protect\subref{subfig:2j_plus_NNpost} \srp and
\protect\subref{subfig:2j_minus_NNpost} \srm are compared with the expected distributions (histograms) from simulated events after the fit (post-fit). In these distributions, the signal contribution is shown stacked on top of contributions from all contributing background processes.
All uncertainties considered in the analysis are included in the
hatched uncertainty band. The correlations induced by the fit are taken into account. The lower panel shows the ratio of data and the prediction; in this panel, the uncertainty is displayed as a grey band.
}
\end{figure}
The correlations induced by the maximum-likelihood fit are taken into account and lead to a
large reduction in the size of the uncertainty band.
The post-fit event yields of the different processes are provided in Table \ref{tab:postfityields}.
\begin{table}[htb!]
\centering
\sisetup{
table-align-uncertainty=true,
separate-uncertainty=true,
}
\renewrobustcmd{\bfseries}{\fontseries{b}\selectfont}
\renewrobustcmd{\boldmath}{}
\hspace*{-1cm}
\begin{tabular}{
c
S[table-format=6.0,detect-weight]@{${}\pm{}$}S[table-format=5.0,detect-weight]
S[table-format=6.0,detect-weight]@{${}\pm{}$}S[table-format=5.0,detect-weight]
}
\toprule
Process &\multicolumn{2}{c}{\srp} &\multicolumn{2}{c}{\srm} \\%
\midrule
$tq$      &                      169000 &  6000  &     150 &   150    \\
$\bar{t}q$ &                         90 &    90  &  109000 &  4000    \\
$tW+\bar{t}W,t\bar{b}+\bar{t}b$ & 51000 &  4000  &   49000 &  4000    \\
$t\bar{t}$ &                     265000 & 14000  &  265000 & 14000    \\
$W$+$b\bar{b}$ &                 198000 & 21000  &  159000 & 17000    \\
$W$+$c(\bar{c})$ &                60000 & 13000  &   49000 & 11000    \\
$Z$+jets, diboson &               21000 &  4000  &   19000 &  4000    \\
Multijet &                       50000  & 10000  &   50000 & 10000  \\\midrule
Total &                          814000 &  2100  &  698800 &  2000    \\
\midrule
Observed &\multicolumn{2}{l}{\SI{814185}{}} &  \multicolumn{2}{l}{\SI{698845}{}}    \\
\bottomrule
\end{tabular}
\vspace*{2mm}
\caption{\label{tab:postfityields}The post-fit event yields in the two SRs.
All uncertainties applied in the analysis are included.
Correlations, including anticorrelations, among the nuisance parameters related to the
uncertainties are taken into account as determined in the maximum-likelihood fit,
leading to a reduction in the size of the uncertainties, in particular for the total prediction.
The event yields of the different
processes as quoted in the table do not add up to the total sum given because of rounding effects.}
\end{table}

\paragraph{Dependence on \(m_t\)} The cross-sections and the ratio \rt are determined at a
fixed value of \(\mtop=172.5\,\GeV\). The mass dependence of the measurements
is determined by repeating the measurement with samples of simulated events produced with different
values of \(\mtop\), namely \(\mtop=171\,\GeV\) and \(\mtop=174\,\GeV\).
The dependence of the resulting cross-sections on \mtop is fitted with a first-order polynomial,
for which the constant term is given by the central value at $\mtop = \SI{172.5}{\GeV}$, namely
\begin{equation}
\sigma(\mtop) = \sigma(\SI{172.5}{\GeV}) + a \cdot \Delta\mtop[\si{\GeV}]\,, \nonumber
\end{equation}
where $\Delta\mtop = \mtop - \SI{172.5}{\GeV}$.
The slopes are fitted to be \(a=\SI{-1.50\pm0.26}{\pb \per \GeV}\) for \sigt,
\(a=\SI{-0.85\pm0.31}{\pb \per \GeV}\) for \sigat, and \(a=\SI{-2.35\pm0.69}{\pb \per \GeV}\) for \sigtat.
For \rt the effect is found to be negligible.
\FloatBarrier


\section{Interpretation of the measurements}

%
The measurements of the \tq production cross-sections presented in \Sect{\ref{sec:results}} are interpreted
in different ways. Predictions based on different PDF sets are compared with the measured \rt value
in \Sect{\ref{sec:rt_pdfs}}. A search for a contribution of a four-quark EFT operator to \tq production
using the \dnn distributions in the SRs is presented in \Sect{\ref{sec:eft}}.
The result of this search yields a confidence interval for the EFT coefficient \CqQ.
In addition, the measurement of \sigtat is used to derive limits on the EFT coefficient \CphiQ.
The corresponding operator \OphiQ has the same Lorentz structure as the \(Wtb\) vertex in the SM,
and thus simply scales the cross-section of \tq production, while kinematic distributions are not
altered by its presence.
The CKM matrix element \(V_{tb}\) is extracted from the measurement of
\(\sigtat\) (see \Sect{\ref{sec:vtb}}). In a more general approach, confidence contours are determined
in the \(\fl\absvtd\)-versus-\(\fl\absvtb\), the \(\fl\absvts\)-versus-\(\fl\absvtb\),
and the \(\fl\absvts\)-versus-\(\fl\absvtd\) planes;
the results are presented in \Sect{\ref{sec:CKM_interpretation}}.

\subsection{Sensitivity of \rt to PDF sets}
\label{sec:rt_pdfs}
With an uncertainty of \(+2.2\%\) / \(-2.1\%\), the measurement of \rt can potentially distinguish
between different PDF sets. Predictions of \rt made with different PDF sets at NNLO, namely
ABMP~\cite{Alekhin:2018pai}, ATLAS~\cite{STDM-2020-32},
CTEQ~\cite{Hou:2019efy}, MSHT~\cite{Bailey:2020ooq},
\nnpdf~\cite{Ball:2014uwa,NNPDF:2021njg}, and PDF4LHC~\cite{PDF4LHCWorkingGroup:2022cjn},
are compared with the measured value in Figure~\ref{fig:rt}.
\begin{figure}[bth!]
\centering
\includegraphics[width=0.75\textwidth]{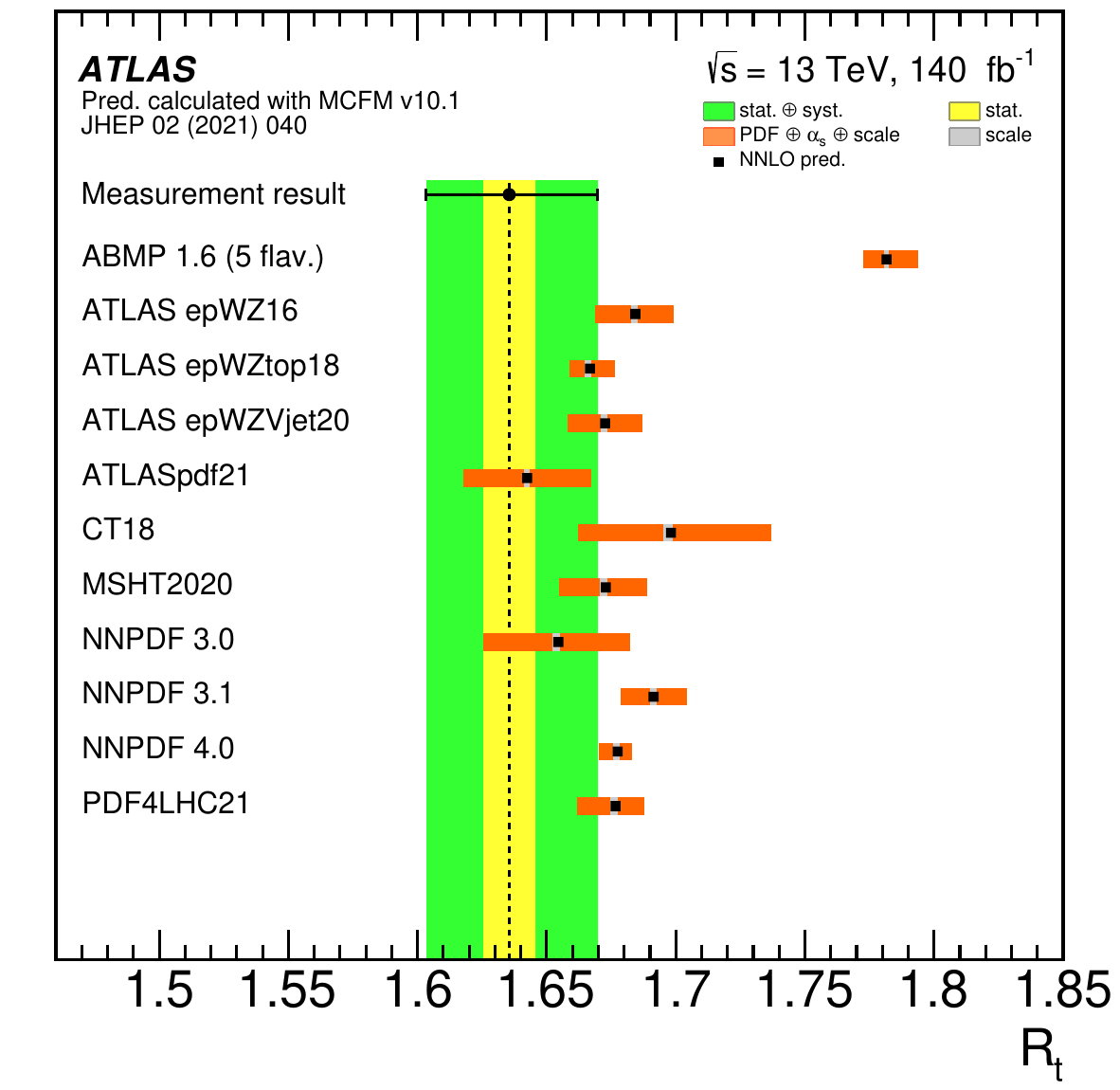}
\caption{\label{fig:rt} The measured value of \rt (dot). The yellow band represents the statistical
uncertainty and the green band represents the total uncertainty of the measurement.
For comparison, the NNLO predictions of MCFM based on different PDF sets are included:
ABMP~\cite{Alekhin:2018pai}, ATLAS~\cite{STDM-2020-32}, CTEQ~\cite{Hou:2019efy},
MSHT~\cite{Bailey:2020ooq}, \nnpdf~\cite{Ball:2014uwa,NNPDF:2021njg}, and
PDF4LHC~\cite{PDF4LHCWorkingGroup:2022cjn}.
The uncertainties in the theoretical predictions include PDF, scale and \alphas uncertainties.}
\end{figure}
The calculations were performed with the \MCFM[10.1] program~\cite{Campbell:2020fhf}.
The differences between the \rt predictions are driven by differences between the \(u\)- and \(d\)-quark PDFs.
The PDFs provided by the different groups differ in the data used, the
value of \alphas assumed, the values of quark masses used, and the treatment of
heavy quarks. The scale uncertainties for the theoretical predictions are included
in Figure~\ref{fig:rt}.
The scale uncertainties are determined %
by varying \(\muR\) and \(\muF\) independently up and down by a factor of two,
whilst never allowing them to differ by a factor greater than two from each other.
The scale uncertainty is defined as the envelope of the six resulting variations.
The uncertainties in the predictions also include the uncertainties provided by the
PDF set under investigation and, where possible, uncertainties in
\alphas.\footnote{There is no functionality implemented to vary \alphas
for ATLAS (epWZ16). For MSHT the strong coupling constant is varied simultaneously
with the PDF eigenvectors rather than independently.}

The prediction of ABMP is incompatible with the measurement of \(\rt\) at the level of approximately
three standard deviations. All other predictions are in agreement with the measured value within
the experimental and theoretical uncertainties.
The predictions of ATLASpdf21 and \nnpdf{3.0} are the closest to the central value of the measurement,
while all other predictions are approximately one standard deviation above.
This is comparable to the difference seen between predictions and the ATLAS measurement at a centre-of-mass energy of
\(8\,\TeV\) in \RunOne of the LHC~\cite{TOPQ-2015-05}.
The slightly higher uncertainty of the ATLASpdf21 prediction compared with those predictions based on other ATLAS PDF sets
is attributed to the usage of a wider range of input data samples and, associated to that,
a modified uncertainty definition.

\subsection{EFT interpretation}
\label{sec:eft}
The Standard Model Effective Field Theory (SMEFT) provides a model-independent framework for
indirect searches for new physics. Within this framework, the SM is regarded as a low-energy
approximation of a more fundamental theory involving interactions at an energy scale $\Lambda$.
The impact of new physics is parameterised by higher-dimensional operators maintaining SM symmetries.
The effective Lagrangian is given by
\[
\mathcal{L}_{\mathrm{eff}}=\mathcal{L}_{\mathrm{SM}}+\sum_{i} \frac{C_{i}}{\Lambda^{2}} O_{i}
+ \mathrm{Hermitian\; conjugate},
\]
where $\mathcal{L}_{\mathrm{SM}}$ is the SM Lagrangian. The $O_i$ are effective dimension-6 operators
and the $C_i$ are the associated Wilson coefficients.
In this EFT interpretation, two operators are considered, the four-quark operator \OqQ
and the operator \OphiQ coupling the third quark generation to the Higgs boson doublet \(\Phi\).

The relevant operators, expressed in the Warsaw basis, are
\begin{eqnarray}
O_{qq}^{1(ijkl)} & = & (\bar{q}_i \gamma^\mu q_j)(\bar{q}_k \gamma_\mu q_l), \nonumber \\
O_{qq}^{3(ijkl)} & = & (\bar{q}_i \gamma^\mu \tau^I q_j)(\bar{q}_k \gamma_\mu \tau^I q_l) \ \ \ \mathrm{and} \nonumber \\
O_{\phi Q}^{3} & = & i(\Phi^+ \tau^I D_\mu \Phi)(\bar{Q} \gamma^\mu \tau^I Q). \nonumber
\end{eqnarray}
The \(q\) denote weak-isospin doublets with $ijkl\in 1,2,3$ as quark generation indices, while $Q$
represents the doublet of the third quark generation.
All contributing four-quark processes depend solely on a linear combination of Wilson coefficients
\begin{equation}
C_{Qq}^{3,1} = \sum_{i=1,2} C_{qq}^{3(ii33)} + \frac{1}{6} C_{qq}^{1(i33i)} - \frac{1}{6} C_{qq}^{3(i33i)},
\nonumber
\end{equation}
therefore the four-quark interaction is fully characterised by \OqQ~\cite{Buckley:2016}.

The operator \OqQ leads to non-SM single top-quark production, as illustrated in
\Fig{\ref{fig:Feynman_EFT_tchan}}, for example via the
process \(b+u\rightarrow t+d\). Top quarks produced in this way feature different angular
distributions and kinematics than those produced via SM processes.
\begin{figure}[!tb]
\centering
\includegraphics[width=0.25\textwidth]{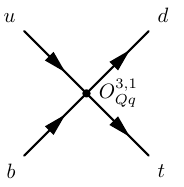}
\caption{Representative LO Feynman diagram of a four-quark contact interaction leading to the
production of a single top quark.}
\label{fig:Feynman_EFT_tchan}
\end{figure}
A non-zero contribution from the operator \OphiQ does not alter the Lorentz structure of the \(Wtb\) vertex,
it merely leads to a rescaling of the vertex strength, and thus changes the total cross-section,
but it does not alter the shape of any kinematic distributions.

Events with single top quarks produced via SM vertices (\tq and \(t\bbar\) production)
and via \OqQ vertices were
generated with \MADGRAPH, as detailed in \Sect{\ref{sec:samples_EFT_CKM}}.
These samples are subjected to the regular analysis chain, including the event selection and
the processing of the NN analysis.
Using the five different EFT samples,
the single top-quark event yield \(\nu_j\) in each bin \(j\) of the \dnn distributions
in \srp and \srm is
parameterised by a polynomial of second degree in the EFT coefficient \(C_{Qq}^{3,1}\).
\begin{equation}%
\nu_j=\nu_{0j}+a_{1j}\frac{C_{Qq}^{3,1}}{\Lambda^2} +
a_{2j}\frac{(C_{Qq}^{3,1})^2}{\Lambda^4}. \nonumber
\end{equation}
The constant term \(\nu_{0j}\) represents \tq and  \(t\bbar\) production in the SM and is normalised
to the SM cross-section predictions of both the processes, as reported in \Sect{\ref{sec:samples_top_quark}}.
The term linear in \(C_{Qq}^{3,1}\) represents the effect of the interference of SM and non-SM amplitudes and
the term proportional to \(\left(C_{Qq}^{3,1}\right)^2\) is entirely due to the four-quark
operator.
Based on a maximum-likelihood scan of the parameter \CqQ the 95\% CL
interval is determined to be
\[ -0\textnormal{.}37 < \CqQ < 0\textnormal{.}06. \]
All SM processes are modelled in the same way as in the cross-section measurements presented
in \Sect{\ref{sec:results}}. The same systematic
uncertainties are applied.
The constraints presented improve the limits set by the ATLAS measurement of the charge asymmetry
in \(\ttbar\) production~\cite{TOPQ-2020-06}, which obtained a confidence interval of
\([-0.70,0.75]\).
The interpretation of cross-section measurements of \(\ttbar Z\) production by
ATLAS~\cite{atlascollaboration2023inclusive} reaches the constraints \([-0.34,0.23]\),
similar to the ones presented above.
Limits on the parameter \CqQ were also set by global EFT
fits that include inputs from measurements by the ATLAS and CMS Collaborations.
The results in \Refn{\cite{Ethier:2021bye}} are based on various measurements of Higgs boson,
diboson and top-quark production processes and lead to a confidence interval for \CqQ
of \([-0.088, 0.166]\) at the 95\% CL when including terms of order \(\Lambda^{-4}\).
A similar approach by a different group of analysers~\cite{Ellis:2020unq} leads to a confidence interval of
\([-0.043, 0.16]\). Using only top-quark measurements for the analysis, the authors of
\Refn{\cite{Brivio:2019ius}} obtain \(\CqQ\in[-0.39, 0.11]\). The comparison to these results of
global EFT analyses demonstrates that the limits on \CqQ obtained from the \tq cross-section
measurements presented in this document are quite competitive. An important difference to appreciate is that
the results of Refs.~\cite{Ethier:2021bye,Ellis:2020unq,Brivio:2019ius} do not account for
reconstruction effects on EFT signal events, while the results presented here are based on
simulated samples that include detector effects.

Since the EFT operator \OphiQ has the same Lorentz structure as the \(Wtb\) vertex in the SM,
and kinematic distributions of \tq and \tbarq events are thus not altered by contributions from
\OphiQ, limits on
the corresponding Wilson coefficient \CphiQ are derived from the measured cross-section \sigtat.
The cross-section \sigtat is calculated for different values of \CphiQ with \MGNLO[2.7.3] using the
SMEFTatNLO-NLO model~\cite{Degrande:2020evl} with the five-flavour scheme and the \NNPDF[3.0nlo] PDF
set, and is obtained as the sum of \sigt and \sigat.
The contribution of the quadratic term in \CphiQ is negligible in the relevant parameter range, and thus
a linear function is fitted to the relative change in \sigtat as a function of \CphiQ relative to its
value at \(\CphiQ=0\), resulting in a slope of \(0.12\pm0.02\).
Based on this parameterisation, the 95\% CL interval of \CphiQ is determined to
be
\begin{equation}\nonumber
-0\textnormal{.}87 < \CphiQ < 1\textnormal{.}42\textnormal{.}
\end{equation}
These constraints improve limits obtained by the interpretation of cross-section
measurements of \(\ttbar Z\) production that yielded the confidence interval
\([-0.95,2.0]\)~\cite{atlascollaboration2023inclusive}. However, the combined interpretation of Higgs boson,
diboson, and top-quark measurements yielded a stronger limit of \([-0.375,0.344]\)~\cite{Ethier:2021bye}.

\subsection{Determination of \(|V_{tb}|\)}
\label{sec:vtb}
Single top-quark production in the $t$-channel proceeds primarily via a $Wtb$ vertex and
the cross-section is proportional to \flvtbsqr.
In the SM, the left-handed form factor $\fl$ is exactly one and the CKM matrix is unitary.
Assuming the unitarity relations, the measured values of other CKM matrix elementes suggest that \absvtb is very close to one.
However, new-physics contributions could alter the value of $\fl$ significantly.
The determination of \flvtb based on single-top-quark
cross-section measurements is independent of assumptions about the number of
quark generations and the unitarity of the CKM matrix.
The only assumptions made are that $\absvtb\gg|V_{td}|,|V_{ts}|$
and that the $Wtb$ interaction is a left-handed weak coupling, as in the SM.

The value of \flvtbsqr is extracted by dividing the measured value of
\(\sigtat\) by the SM expectation of
\(214.2\pm3.4(\mathrm{scale + PDF})\pm0.6(\Delta m_t)\,\)pb~\cite{Campbell:2020fhf}.
When calculating \flvtbsqr, the experimental and theoretical
uncertainties are added in quadrature.
The uncertainty in \(m_t\) is also considered, assuming $\Delta m_t = \SI{\pm 0.33}{\GeV}$
based on a combination of \RunOne measurements of \(m_t\) by the ATLAS and CMS
Collaborations~\cite{TOPQ-2019-13}.
The result obtained is
\[\flvtb = 1.015 \pm 0.031, \]
improving the precision by 30\% compared with the combination of \RunOne
measurements by ATLAS and CMS~\cite{TOPQ-2017-16}.
The Particle Data Group combined all available measurements performed at the Tevatron and the
LHC to \(1.014\pm0.029\)~\cite{Workman:2022ynf}.

Restricting the range of \absvtb to the interval $[0, 1]$ and setting \fl=1, as required by
the SM, a lower limit on \absvtb is extracted:
$\absvtb>0.95$ at the \SI{95}{\%} CL.
In the Bayesian-style limit computation, it is assumed that
the likelihood curve of $|V_{tb}|^2$ is a Gaussian function, centered at the measured
value. A flat prior in $|V_{tb}|^2$ is applied, being one in the interval $[0, 1]$ and
zero otherwise.

\subsection{Generalised CKM interpretation}
\label{sec:CKM_interpretation}
The interpretation of the \tq cross-section measurements presented in \Sect{\ref{sec:vtb}} neglects the
contributions due to \(Wts\) and \(Wtd\) vertices. In a more general approach, this caveat is avoided.
Nine contributions to \tq production are considered, differing in the combination of \(Wtq\) vertices
for top-quark production and decay with \(q\in\{d, s, b\} \).
In \(t\tbar\) production, \(Wtq\) vertices occur for the top-quark and top-antiquark decays.
Again, nine different combinations of vertices are considered, thus treating the most important
background process at the same level of modelling as the \tq signal process.
Including the effect of \(Wts\) and \(Wtd\) vertices for \(t\tbar\) production improves the
sensitivity of the measurement to \absvts and \absvtd by approximately 20\%.
The effects of \(Wts\) and \(Wtd\) vertices on the event yields of \(tW\) and \(t\bbar\) production
are neglected, since the corresponding event yields for these processes are much smaller than
the yields for \tq and \(t\tbar\) production.
Three different fit scenarios are investigated. In each scenario, two \(V_{tq}\) matrix elements
are considered to be free parameters, while the third parameter is fixed to be either \(0\) or \(1\):
\begin{center}
\begin{tabular}{ll}
Scenario 1 & \(\absvtb\neq0\), \(\absvtd\neq0\) and \(\absvts=0\), \\
Scenario 2 & \(\absvtb\neq0\), \(\absvts\neq0\) and \(\absvtd=0\), \\
Scenario 3 & \(\absvtd\neq0\), \(\absvts\neq0\) and \(\fl\absvtb=1\).
\end{tabular}
\end{center}
The form factor \(\fl\) is non-zero in all scenarios and is also allowed to be greater than one.
For each scenario, a maximum-likelihood scan of the two non-zero CKM matrix elements is performed. As a result,
confidence contours are determined at the 95\% CL in the
\(\fl\absvtd\)-versus-\(\fl\absvtb\), the \(\fl\absvts\)-versus-\(\fl\absvtb\),
and the \(\fl\absvts\)-versus-\(\fl\absvtd\) planes.
These contours are shown in \Fig{\ref{fig:LLscan_CKM}}.
\begin{figure}[!t!bh]
\centering
\subfloat[][]{
\includegraphics[width=0.46\textwidth]{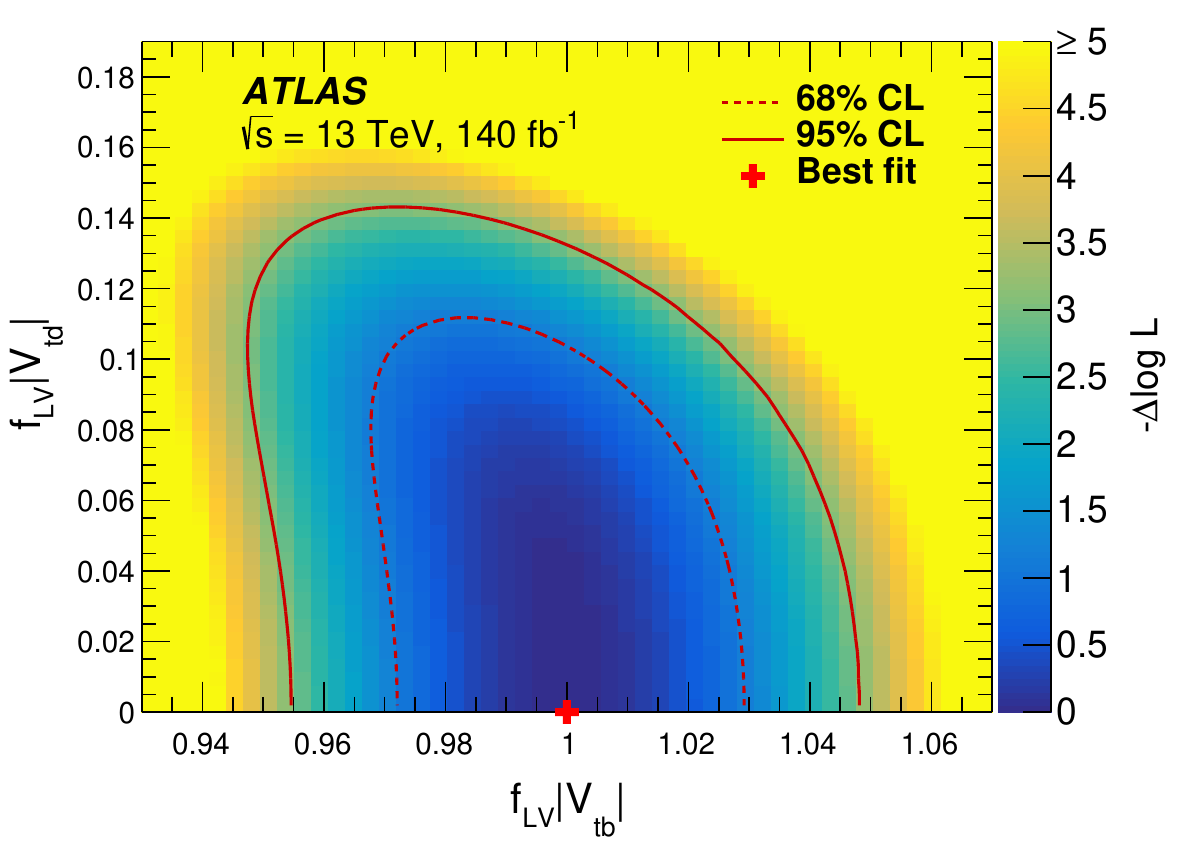}
\label{subfig:LLscan_VtbVtd}
}
\quad
\subfloat[][]{
\includegraphics[width=0.46\textwidth]{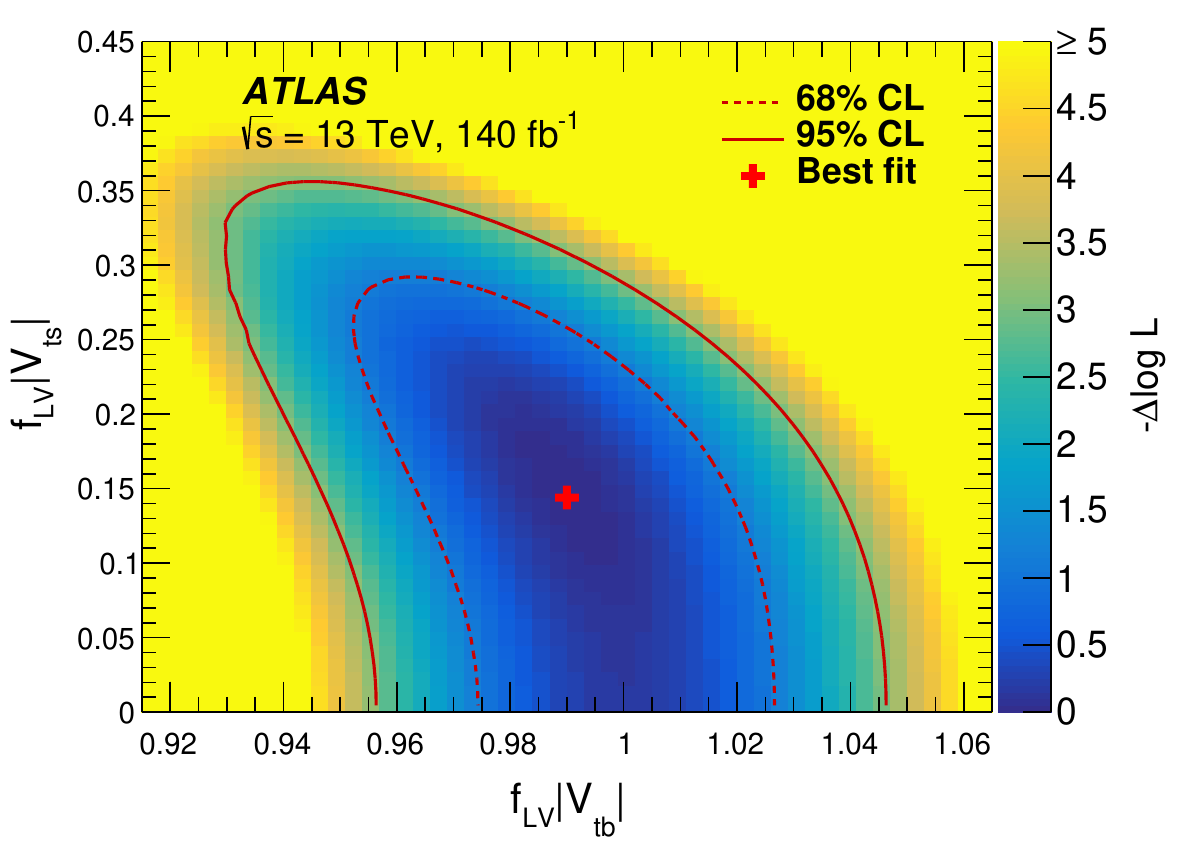}
\label{subfig:LLscan_VtbVts}
} \\
\subfloat[][]{
\includegraphics[width=0.46\textwidth]{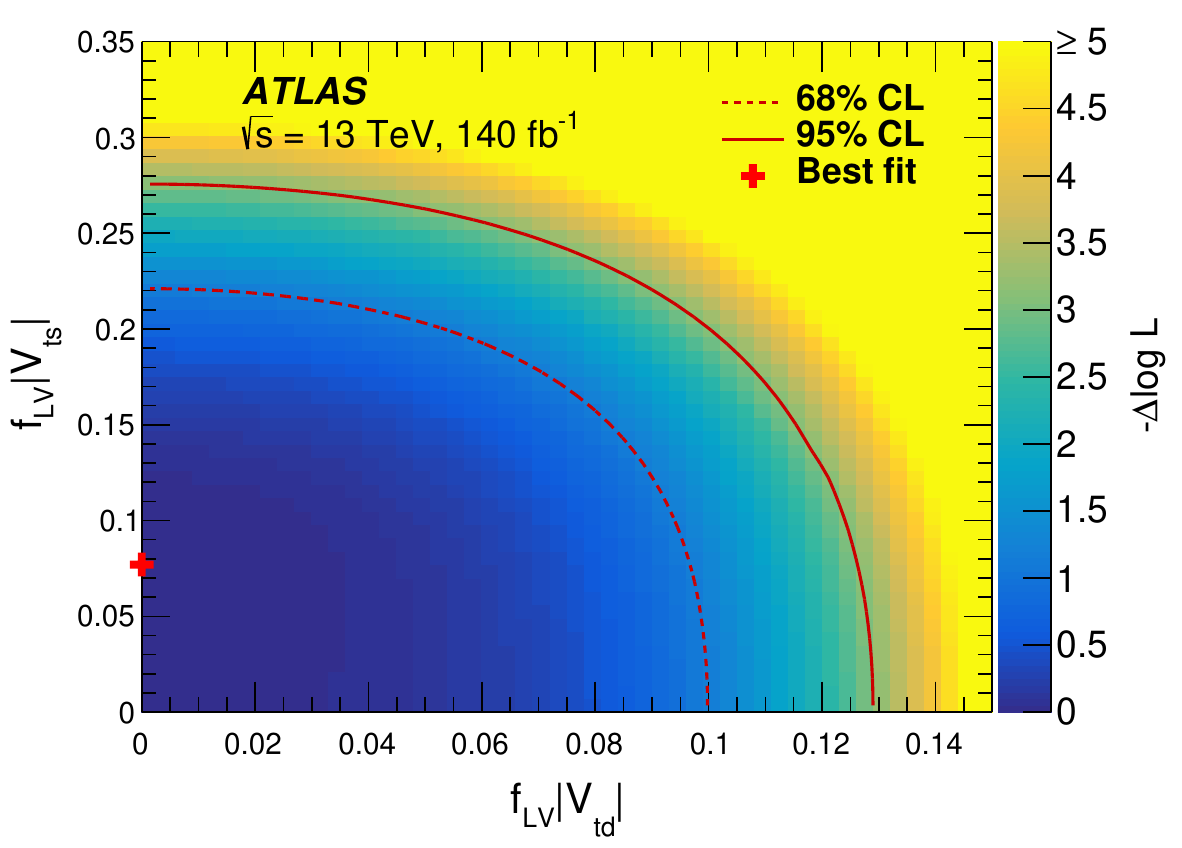}
\label{subfig:LLscan_VtdVts}
}
\caption{\label{fig:LLscan_CKM}Confidence contours obtained from maximum-likelihood scans in
\protect\subref{subfig:LLscan_VtbVtd} the \(\fl\absvtd\)-versus-\(\fl\absvtb\) plane,
\protect\subref{subfig:LLscan_VtbVts} the \(\fl\absvts\)-versus-\(\fl\absvtb\) plane, and
\protect\subref{subfig:LLscan_VtdVts} the \(\fl\absvts\)-versus-\(\fl\absvtd\) plane.
Contours at the 68\% and 95\% confidence levels are shown.
The two-dimensional histogram contains the values of the difference of the
log-likelihood function at a certain point in the plane to the minimum of the log-likelihood function
indicated by the red cross.
}
\end{figure}
The parameter \(\fl\absvtb\) is constrained at the 95\% CL
to a range between \(0.94\) and \(1.04\),
and \(\fl\absvtd\) and \(\fl\absvts\) are constrained to be \(<0.13\) and \(<0.31\), respectively. The constraint on \(\absvtd\) is stronger compared with the one on \(\absvts\) because the
\(d\)-quark is a valence quark of the proton, while \(s\)-quarks appear as sea-quarks only.
The interpretation uses the nominal simulation-to-data corrections for the efficiency of tagging
\(b\)-quark jets, which was determined with \(t\tbar\) events in the dilepton channel assuming
\(\absvtb=1\) and thus \(\mathcal{B}(t\rightarrow Wb)=100\%\).
This assumption is increasingly violated when moving to large values of \absvtd and \absvts
and thus constitutes a caveat of the generalised CKM interpretation presented here.
A similar model-independent measurement of the modulus of the CKM matrix elements
\absvtb, \absvtd, and \absvts was performed by the CMS Collaboration and yielded
\(\absvtb=0.988\pm0.024\) and \(\absvtd^2+\absvts^2=0.06\pm0.06\)~\cite{CMS-TOP-17-012}.

The CKM matrix elements \absvtd and \absvts are determined very precisely using measurements
of the mass differences \(\Delta m_d\) and \(\Delta m_s\) of the mass eigenstates of
\(B^0_d\) and \(B^0_s\) mesons~\cite{Workman:2022ynf}. However, these determinations are based
on \(B\)--\(\bar{B}\) meson oscillations that are induced by box diagrams with top quarks,
use lattice QCD results, and neglect corrections suppressed by \(\absvtb-1\), and thus
introduce a certain level of model dependence that is reduced in the studies presented here for
tree-level processes, namely single top-quark production and decay. In addition, the top-quark
processes give access to a much higher energy scale.


\FloatBarrier

\section{Conclusions}
The production of single top quarks and top antiquarks via the \(t\)-channel exchange of a
virtual \(W\) boson is measured in proton--proton collisions at the LHC at a centre-of-mass energy
of \(13\,\TeV\), using the full \RunTwo data sample of 140~\ifb recorded with the ATLAS detector.
Events are selected with either one isolated electron or muon, high \MET, and exactly
two hadronic jets with high \pt. Exactly one of these jets is required to be \btagged.
An artificial NN is used to construct a discriminant that separates signal and background events.
The distributions of the discriminant are used in profile maximum-likelihood fits to determine the signal yields.

The total cross-sections are determined to be \(\sigt=137^{+8}_{-8}\,\mathrm{pb}\)
and \(\sigat=84^{+6}_{-5}\,\mathrm{pb}\) for top-quark and top-antiquark production,
respectively.
The combined cross-section is found to be \(\sigtat=221^{+13}_{-13}\,\mathrm{pb}\) and the
cross-section ratio is \(\rt=\sigt/\sigat=1.636^{+0.036}_{-0.034}\).
The predictions made at \NNLO in perturbation theory are in good agreement with the
measured cross-sections, which reach greater precision than
previous measurements by the ATLAS and CMS Collaborations with partial \RunTwo data samples
at \sqrts. The relative precision of the measurements presented also
surpasses the precision reached in ATLAS \RunOne measurements at
\(\sqrt{s}=7\,\TeV\) and \(\sqrt{s}=8\,\TeV\).
The new results are thus the most precise measurements of \(tq\) and \(\tbar q\) production to date.

The predictions using various sets of PDFs are compared with the measured value of \rt,
demonstrating the potential of further constraining the functions if the
measurement is included into future fits.
The measurements of \sigt, \sigat, and \sigtat are interpreted in an EFT approach,
setting limits at the 95\% CL on the strength of the four-quark operator
\OqQ and the operator \CphiQ:
\(-0\textnormal{.}37 < \CqQ < 0\textnormal{.}06\) and
\(-0\textnormal{.}87 < \CphiQ < 1\textnormal{.}42\), respectively.
The measured value of \sigtat is further used to derive the constraint \(\absvtb>0.95\) at the 95\%
CL and determine \(\flvtb = 1.015 \pm 0.031, \)
improving by 30\% the determination of this quantity based on a combination of \RunOne
measurements by ATLAS and CMS.
In a more general approach, confidence contours are determined
in the \(\fl\absvtd\)-versus-\(\fl\absvtb\), the \(\fl\absvts\)-versus-\(\fl\absvtb\), and the
\(\fl\absvts\)-versus-\(\fl\absvtd\) planes.
The parameter \(\fl\absvtb\) is constrained at the 95\% CL
to a range between \(0.94\) and \(1.04\),
and \(\fl\absvtd\) and \(\fl\absvts\) are constrained to be \(<0.13\) and \(<0.31\),
respectively.

\section*{Acknowledgements}

%
%

%
%

We thank CERN for the very successful operation of the LHC and its injectors, as well as the support staff at
CERN and at our institutions worldwide without whom ATLAS could not be operated efficiently.

The crucial computing support from all WLCG partners is acknowledged gratefully, in particular from CERN, the ATLAS Tier-1 facilities at TRIUMF/SFU (Canada), NDGF (Denmark, Norway, Sweden), CC-IN2P3 (France), KIT/GridKA (Germany), INFN-CNAF (Italy), NL-T1 (Netherlands), PIC (Spain), RAL (UK) and BNL (USA), the Tier-2 facilities worldwide and large non-WLCG resource providers. Major contributors of computing resources are listed in Ref.~\cite{ATL-SOFT-PUB-2023-001}.

We gratefully acknowledge the support of ANPCyT, Argentina; YerPhI, Armenia; ARC, Australia; BMWFW and FWF, Austria; ANAS, Azerbaijan; CNPq and FAPESP, Brazil; NSERC, NRC and CFI, Canada; CERN; ANID, Chile; CAS, MOST and NSFC, China; Minciencias, Colombia; MEYS CR, Czech Republic; DNRF and DNSRC, Denmark; IN2P3-CNRS and CEA-DRF/IRFU, France; SRNSFG, Georgia; BMBF, HGF and MPG, Germany; GSRI, Greece; RGC and Hong Kong SAR, China; ISF and Benoziyo Center, Israel; INFN, Italy; MEXT and JSPS, Japan; CNRST, Morocco; NWO, Netherlands; RCN, Norway; MEiN, Poland; FCT, Portugal; MNE/IFA, Romania; MESTD, Serbia; MSSR, Slovakia; ARRS and MIZ\v{S}, Slovenia; DSI/NRF, South Africa; MICINN, Spain; SRC and Wallenberg Foundation, Sweden; SERI, SNSF and Cantons of Bern and Geneva, Switzerland; MOST, Taipei; TENMAK, T\"urkiye; STFC, United Kingdom; DOE and NSF, United States of America.

Individual groups and members have received support from BCKDF, CANARIE, CRC and DRAC, Canada; CERN-CZ, PRIMUS 21/SCI/017 and UNCE SCI/013, Czech Republic; COST, ERC, ERDF, Horizon 2020, ICSC-NextGenerationEU and Marie Sk{\l}odowska-Curie Actions, European Union; Investissements d'Avenir Labex, Investissements d'Avenir Idex and ANR, France; DFG and AvH Foundation, Germany; Herakleitos, Thales and Aristeia programmes co-financed by EU-ESF and the Greek NSRF, Greece; BSF-NSF and MINERVA, Israel; Norwegian Financial Mechanism 2014-2021, Norway; NCN and NAWA, Poland; La Caixa Banking Foundation, CERCA Programme Generalitat de Catalunya and PROMETEO and GenT Programmes Generalitat Valenciana, Spain; G\"{o}ran Gustafssons Stiftelse, Sweden; The Royal Society and Leverhulme Trust, United Kingdom.

In addition, individual members wish to acknowledge support from Chile: Agencia Nacional de Investigaci\'on y Desarrollo (FONDECYT 1190886, FONDECYT 1210400, FONDECYT 1230812, FONDECYT 1230987); China: National Natural Science Foundation of China (NSFC - 12175119, NSFC 12275265, NSFC-12075060); Czech Republic: PRIMUS Research Programme (PRIMUS/21/SCI/017); European Union: European Research Council (ERC - 948254), Horizon 2020 Framework Programme (MUCCA - CHIST-ERA-19-XAI-00), European Union, Future Artificial Intelligence Research (FAIR-NextGenerationEU PE00000013), Italian Center for High Performance Computing, Big Data and Quantum Computing (ICSC, NextGenerationEU), Marie Sklodowska-Curie Actions (EU H2020 MSC IF GRANT NO 101033496); France: Agence Nationale de la Recherche (ANR-20-CE31-0013, ANR-21-CE31-0013, ANR-21-CE31-0022), Investissements d'Avenir Idex (ANR-11-LABX-0012), Investissements d'Avenir Labex (ANR-11-LABX-0012); Germany: Baden-Württemberg Stiftung (BW Stiftung-Postdoc Eliteprogramme), Deutsche Forschungsgemeinschaft (DFG - CR 312/5-1); Italy: Istituto Nazionale di Fisica Nucleare (FELLINI G.A. n. 754496, ICSC, NextGenerationEU); Japan: Japan Society for the Promotion of Science (JSPS KAKENHI JP21H05085, JSPS KAKENHI JP22H01227, JSPS KAKENHI JP22H04944); Netherlands: Netherlands Organisation for Scientific Research (NWO Veni 2020 - VI.Veni.202.179); Norway: Research Council of Norway (RCN-314472); Poland: Polish National Agency for Academic Exchange (PPN/PPO/2020/1/00002/U/00001), Polish National Science Centre (NCN 2021/42/E/ST2/00350, NCN UMO-2019/34/E/ST2/00393, UMO-2020/37/B/ST2/01043, UMO-2021/40/C/ST2/00187); Slovenia: Slovenian Research Agency (ARIS grant J1-3010); Spain: BBVA Foundation (LEO22-1-603), Generalitat Valenciana (Artemisa, FEDER, IDIFEDER/2018/048), La Caixa Banking Foundation (LCF/BQ/PI20/11760025), Ministry of Science and Innovation (MCIN \& NextGenEU -PCI2022-135018-2, MICIN \& FEDER -PID2021-125273NB, RYC2019-028510-I, RYC2020-030254-I, RYC2021-031273-I, RYC2022-038164-I), PROMETEO and GenT Programmes Generalitat Valenciana (CIDEGENT/2019/023, CIDEGENT/2019/027); Sweden: Swedish Research Council (VR 2018-00482, VR 2022-03845, VR 2022-04683, VR grant 2021-03651), Knut and Alice Wallenberg Foundation (KAW 2017.0100, KAW 2018.0157, KAW 2018.0458, KAW 2019.0447); Switzerland: Swiss National Science Foundation (SNSF - PCEFP2\_194658); United Kingdom: Leverhulme Trust (Leverhulme Trust RPG-2020-004); United States of America: Neubauer Family Foundation.

%
%


%
%
%
%
%
%

%
\clearpage
\appendix
\part*{Additional plots and interpretations}
\addcontentsline{toc}{part}{Additional plots and interpretations}

%
\Fig{\ref{fig:pie_charts}} illustrates the fractions of selected events in the two SRs for
the different scattering processes based on the post-fit event yields reported in
\Tab{\ref{tab:postfityields}}.
\begin{figure}[!h]
\centering
\subfloat[][]{
\includegraphics[width=0.47\textwidth]{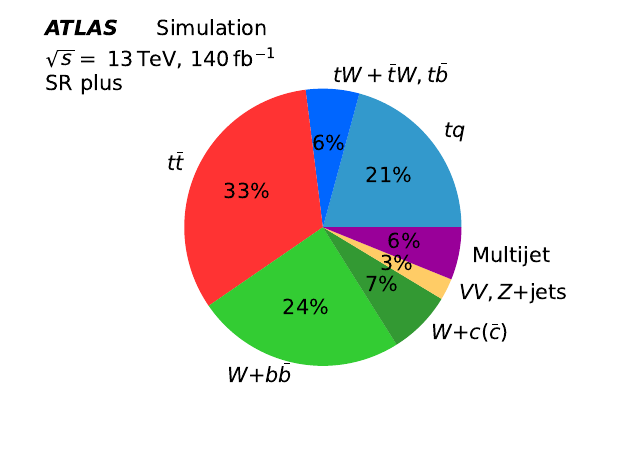}
\label{subfig:pie_SRplus}
}
\quad
\subfloat[][]{
\includegraphics[width=0.47\textwidth]{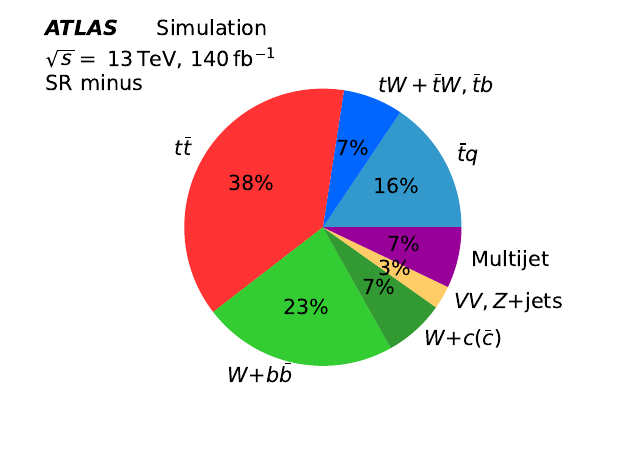}
\label{subfig:pie_SRminus}
}
\caption{\label{fig:pie_charts}Pie chart of the composition of
\protect\subref{subfig:pie_SRplus} the \srp and
\protect\subref{subfig:pie_SRminus} the \srm in terms of the different scattering
processes. The fractions are based on the post-fit event yields.}
\end{figure}

The rate of the multijet background is determined by including CRs enriched in this
background in the maximum-likelihood fit.
\Fig{\ref{fig:CRmuon_DeltaPhi}} shows the post-fit distributions of the variable \(\Delta\phi(\vecpmiss, \mu)\)
in the \crmup and \crmum.
\begin{figure}[!h]
\centering
\subfloat[][]{
\includegraphics[width=0.47\textwidth]{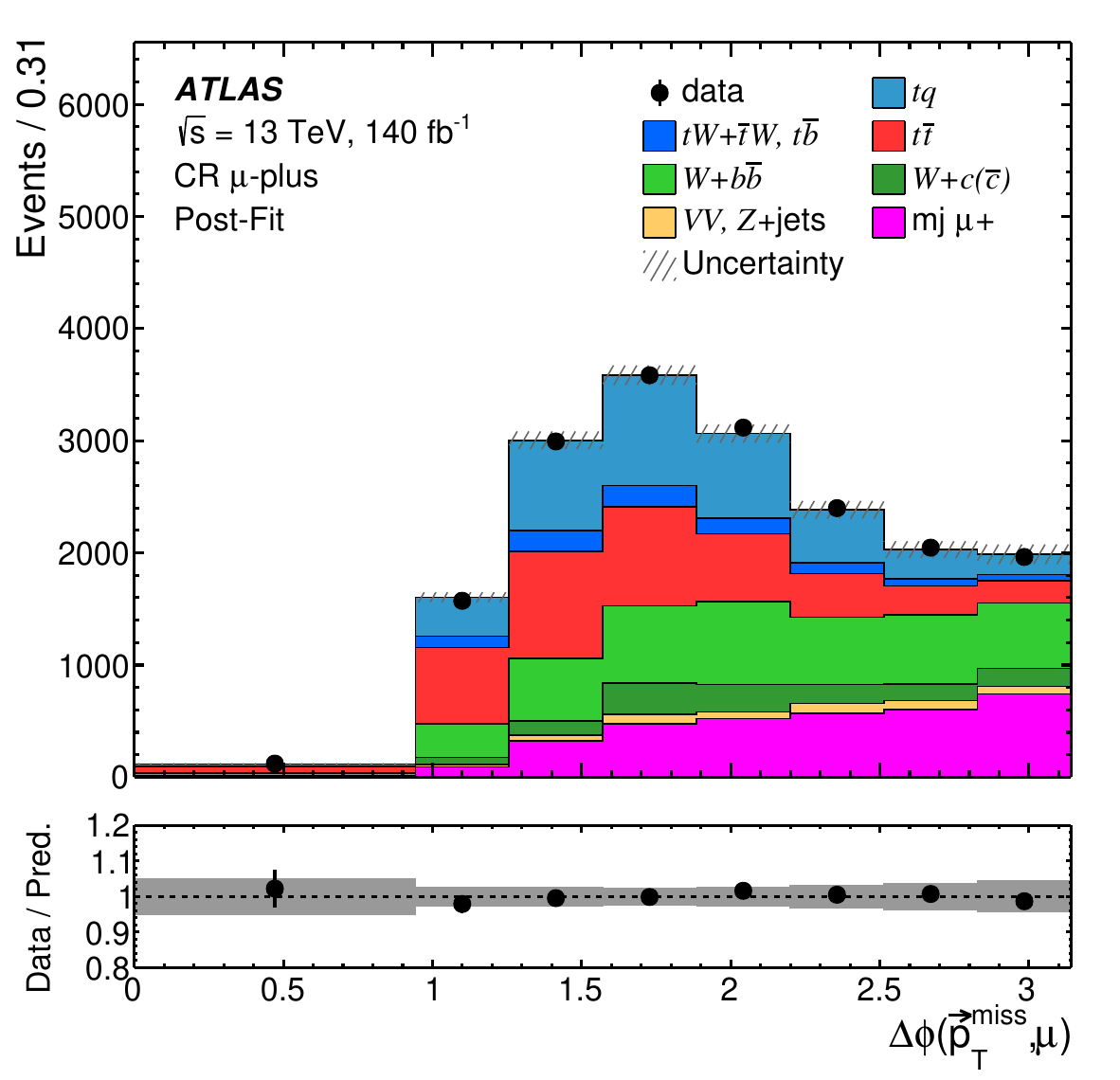}
\label{subfig:CRmuon_plus}
}
\quad
\subfloat[][]{
\includegraphics[width=0.47\textwidth]{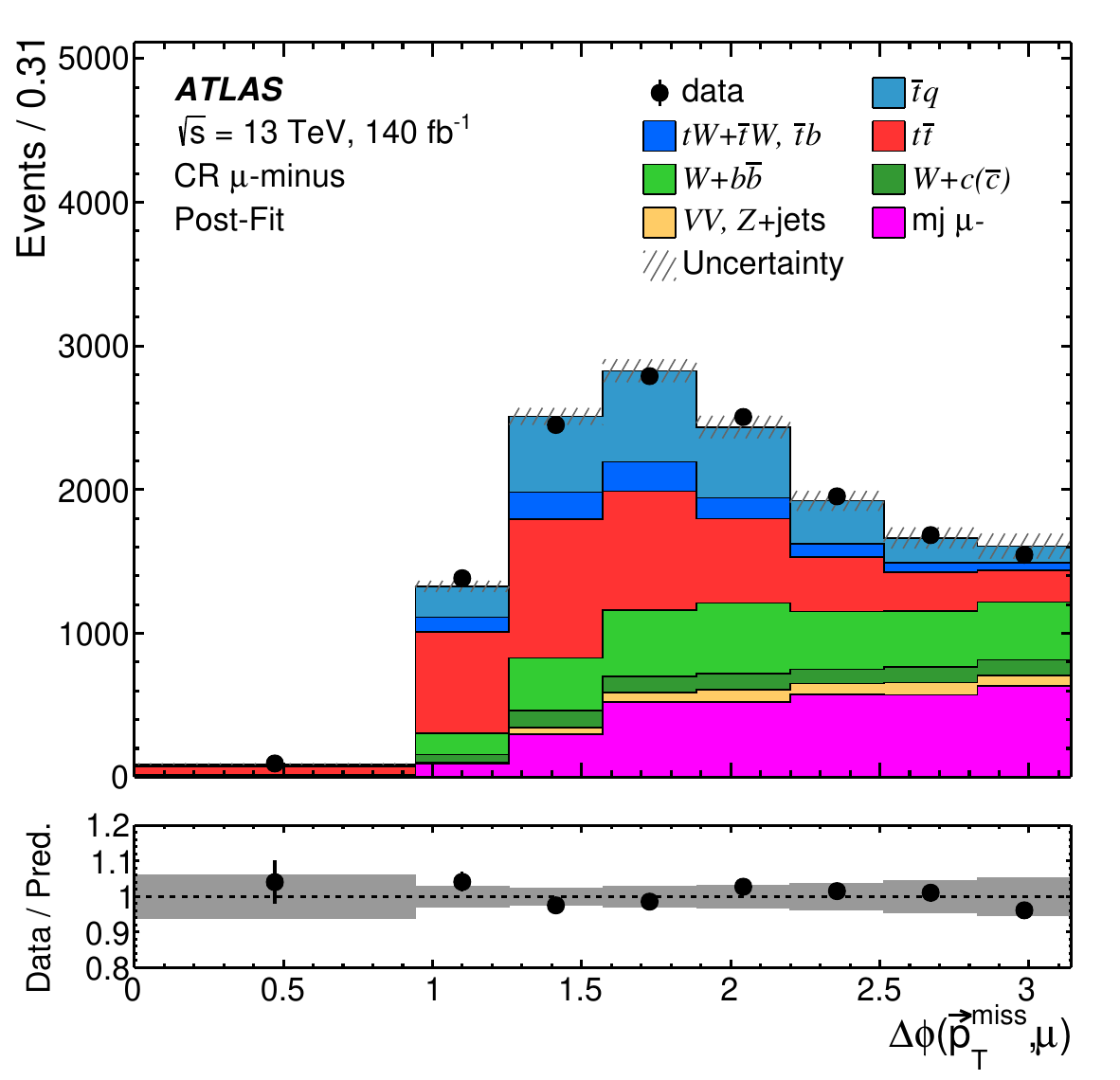}
\label{subfig:CRmuon_minus}
}
\caption{\label{fig:CRmuon_DeltaPhi}Distributions of the variable \(\Delta\phi(\vecpmiss, \mu)\)
in \protect\subref{subfig:CRmuon_plus} the \crmup and \protect\subref{subfig:CRmuon_minus}
the \crmum after the maximum-likelihood fit is performed (post-fit).
In these distributions, the signal contribution is shown stacked on top of contributions from
all contributing background processes.
All uncertainties considered in the analysis are included in the hatched uncertainty band.
The correlations induced by the fit are taken into account.
The lower panel shows the ratio of data and the prediction;
in this panel, the uncertainty is displayed as a grey band.}
\end{figure}

\Fig{\ref{fig:xs_pdfs}} compares NNLO predictions obtained with different PDF sets with
the measured values of \sigt and \sigat. The PDF sets used are
ABMP~\cite{Alekhin:2018pai}, ATLAS~\cite{STDM-2020-32},
CTEQ~\cite{Hou:2019efy}, MSHT~\cite{Bailey:2020ooq},
\nnpdf~\cite{Ball:2014uwa,NNPDF:2021njg}, and PDF4LHC~\cite{PDF4LHCWorkingGroup:2022cjn}.
All predictions agree with the measurements within the uncertainties.
\begin{figure}[!h]
\centering
\subfloat[][]{
\includegraphics[width=0.47\textwidth]{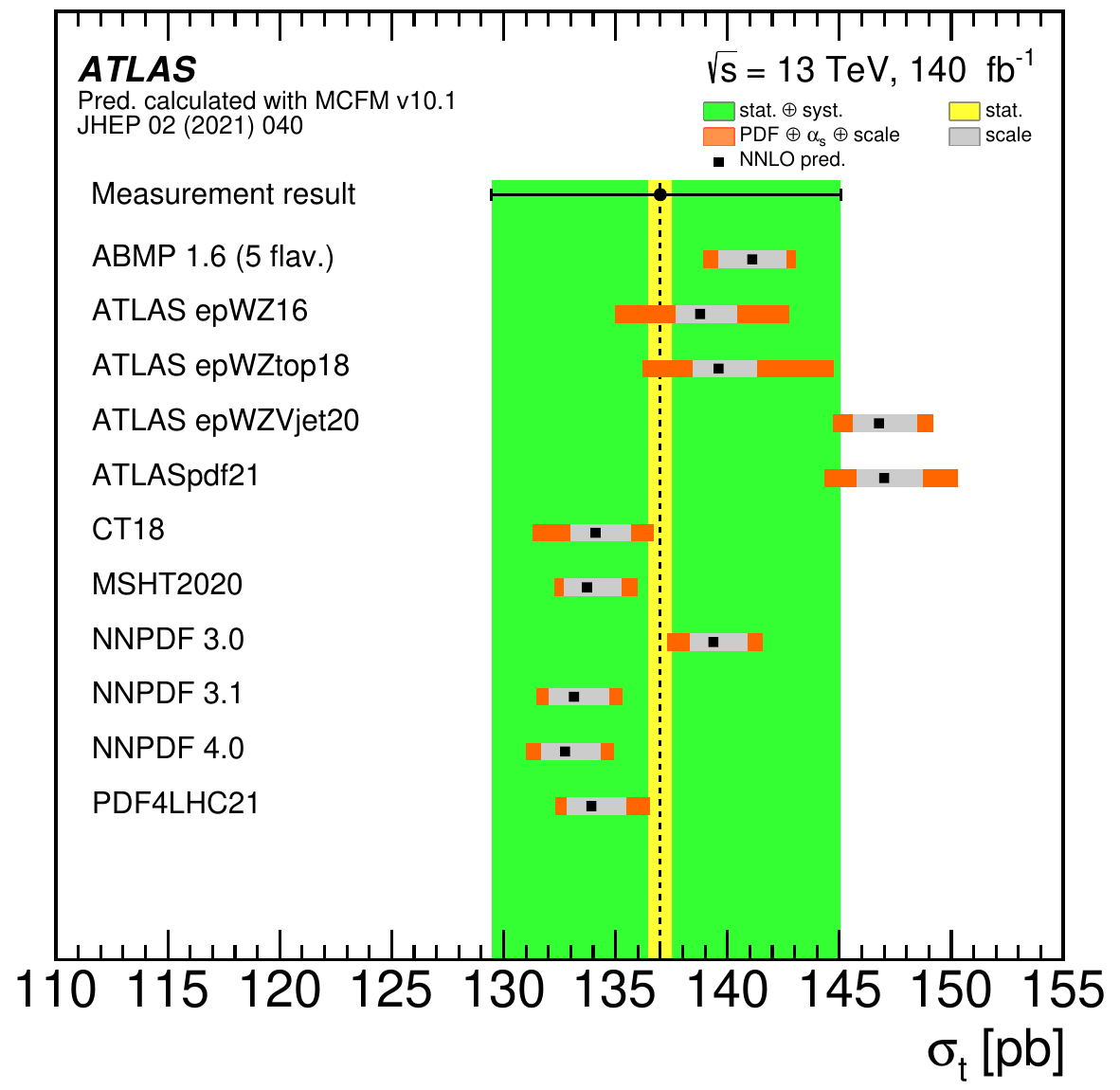}
\label{subfig:sigtopquark_pdf}
}
\quad
\subfloat[][]{
\includegraphics[width=0.47\textwidth]{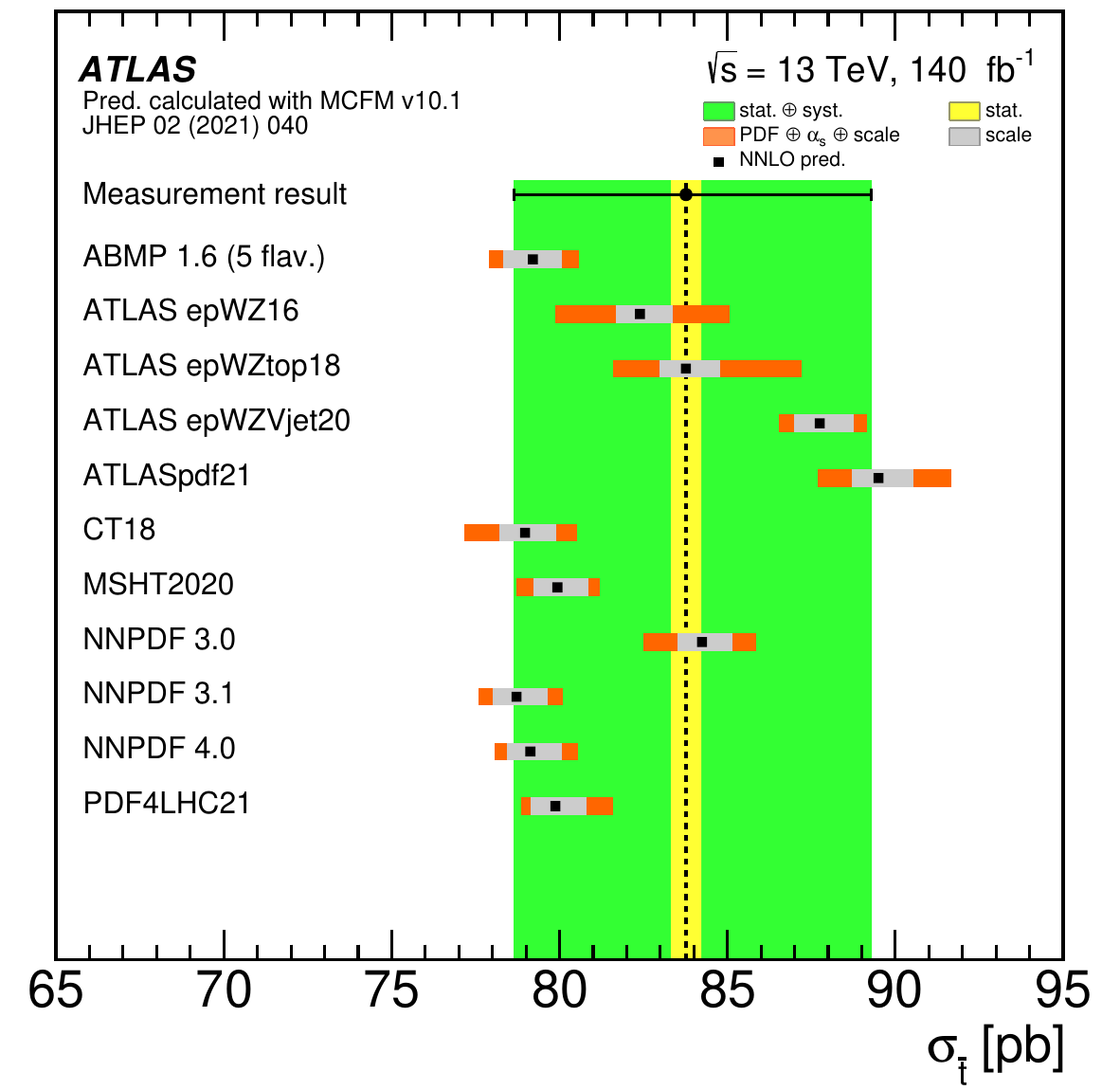}
\label{subfig:sigtopantiquark_pdf}
}
\caption{\label{fig:xs_pdfs}Comparison of NNLO predictions based on several different
PDF sets with the measured values (dots) of
\protect\subref{subfig:sigtopquark_pdf} \sigt and
\protect\subref{subfig:sigtopantiquark_pdf} \sigat. The yellow band represents the statistical
uncertainty and the green band the total uncertainty. The
uncertainties in the theoretical predictions include PDF, scale and
\alphas uncertainties.}
\end{figure}

A 95\% CL interval is determined for the EFT coefficient \(\CqQ\)
by scanning the likelihood function relative to this parameter.
\Fig{\ref{fig:LHscan}} shows the difference between the natural logarithm of the
likelihood function relative to its minimum as a function of \(\CqQ\).
\begin{figure}[!t!hb]
\centering
\includegraphics[width=0.6\textwidth]{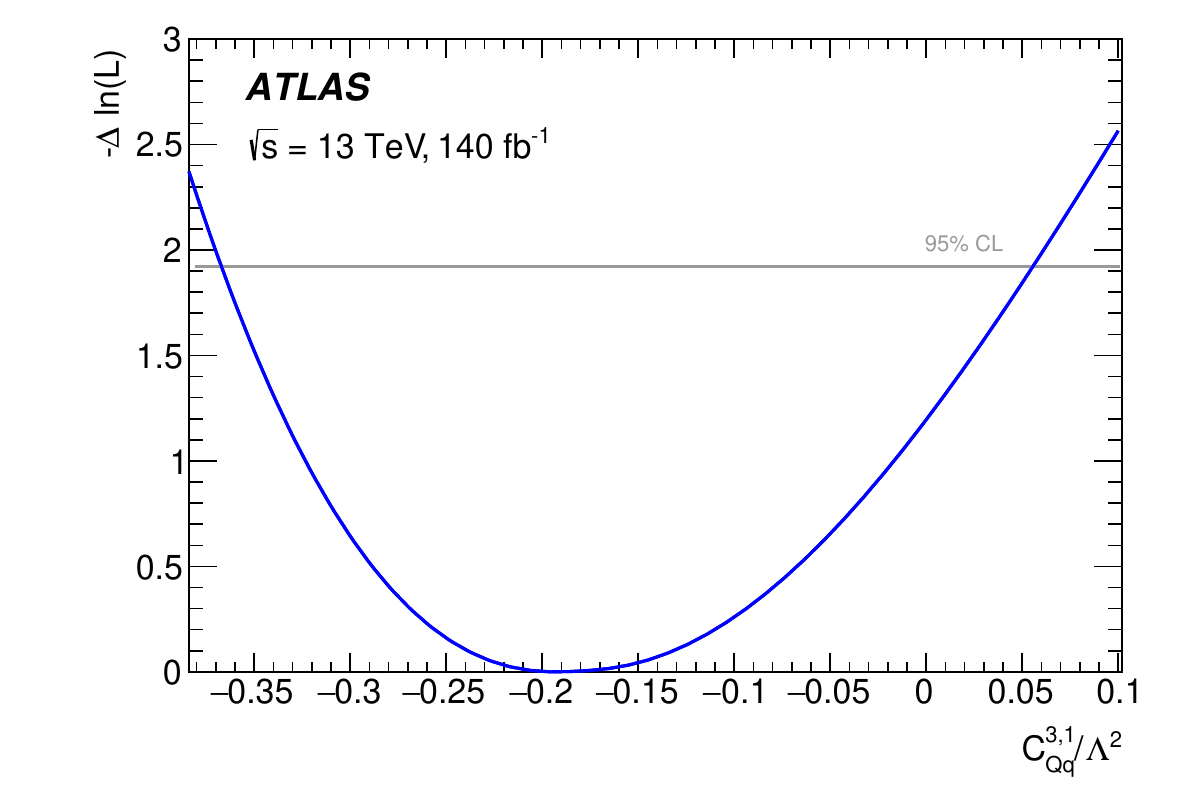}
\caption{Likelihood scan of the EFT coefficient \(\CqQ\).}
\label{fig:LHscan}
\end{figure}


\FloatBarrier
\printbibliography
\addcontentsline{toc}{part}{References}

\clearpage
\input{atlas_authlist}

\end{document}

%% file: atlas_authlist.tex
 
\begin{flushleft}
\hypersetup{urlcolor=black}
{\Large The ATLAS Collaboration}

\bigskip

\AtlasOrcid[0000-0002-6665-4934]{G.~Aad}$^\textrm{\scriptsize 102}$,
\AtlasOrcid[0000-0002-5888-2734]{B.~Abbott}$^\textrm{\scriptsize 120}$,
\AtlasOrcid[0000-0002-1002-1652]{K.~Abeling}$^\textrm{\scriptsize 55}$,
\AtlasOrcid[0000-0001-5763-2760]{N.J.~Abicht}$^\textrm{\scriptsize 49}$,
\AtlasOrcid[0000-0002-8496-9294]{S.H.~Abidi}$^\textrm{\scriptsize 29}$,
\AtlasOrcid[0000-0002-9987-2292]{A.~Aboulhorma}$^\textrm{\scriptsize 35e}$,
\AtlasOrcid[0000-0001-5329-6640]{H.~Abramowicz}$^\textrm{\scriptsize 151}$,
\AtlasOrcid[0000-0002-1599-2896]{H.~Abreu}$^\textrm{\scriptsize 150}$,
\AtlasOrcid[0000-0003-0403-3697]{Y.~Abulaiti}$^\textrm{\scriptsize 117}$,
\AtlasOrcid[0000-0003-0762-7204]{A.C.~Abusleme~Hoffman}$^\textrm{\scriptsize 137a}$
% ...
\bigskip
\\

% ...
$^{29}$Physics Department, Brookhaven National Laboratory, Upton NY; United States of America.\\
% ...
$^{35}$$^{(a)}$Facult\'e des Sciences Ain Chock, R\'eseau Universitaire de Physique des Hautes Energies - Universit\'e Hassan II, Casablanca;$^{(b)}$Facult\'{e} des Sciences, Universit\'{e} Ibn-Tofail, K\'{e}nitra;$^{(c)}$Facult\'e des Sciences Semlalia, Universit\'e Cadi Ayyad, LPHEA-Marrakech;$^{(d)}$LPMR, Facult\'e des Sciences, Universit\'e Mohamed Premier, Oujda;$^{(e)}$Facult\'e des sciences, Universit\'e Mohammed V, Rabat;$^{(f)}$Institute of Applied Physics, Mohammed VI Polytechnic University, Ben Guerir; Morocco.\\
% ...
$^{49}$Fakult\"{a}t Physik , Technische Universit{\"a}t Dortmund, Dortmund; Germany.\\
% ...
$^{55}$II. Physikalisches Institut, Georg-August-Universit\"{a}t G\"ottingen, G\"ottingen; Germany.\\
% ...
$^{102}$CPPM, Aix-Marseille Universit\'e, CNRS/IN2P3, Marseille; France.\\
% ...
$^{117}$Department of Physics, New York University, New York NY; United States of America.\\
% ...
$^{120}$Homer L. Dodge Department of Physics and Astronomy, University of Oklahoma, Norman OK; United States of America.\\
% ...
$^{137}$$^{(a)}$Departamento de F\'isica, Pontificia Universidad Cat\'olica de Chile, Santiago;$^{(b)}$Millennium Institute for Subatomic physics at high energy frontier (SAPHIR), Santiago;$^{(c)}$Instituto de Investigaci\'on Multidisciplinario en Ciencia y Tecnolog\'ia, y Departamento de F\'isica, Universidad de La Serena;$^{(d)}$Universidad Andres Bello, Department of Physics, Santiago;$^{(e)}$Instituto de Alta Investigaci\'on, Universidad de Tarapac\'a, Arica;$^{(f)}$Departamento de F\'isica, Universidad T\'ecnica Federico Santa Mar\'ia, Valpara\'iso; Chile.\\
% ...
$^{150}$Department of Physics, Technion, Israel Institute of Technology, Haifa; Israel.\\
$^{151}$Raymond and Beverly Sackler School of Physics and Astronomy, Tel Aviv University, Tel Aviv; Israel.\\
% ...
$^{a}$ Also Affiliated with an institute covered by a cooperation agreement with CERN.\\
$^{b}$ Also at An-Najah National University, Nablus; Palestine.\\
$^{c}$ Also at Borough of Manhattan Community College, City University of New York, New York NY; United States of America.\\
$^{d}$ Also at Center for High Energy Physics, Peking University; China.\\
$^{e}$ Also at Center for Interdisciplinary Research and Innovation (CIRI-AUTH), Thessaloniki; Greece.\\
$^{f}$ Also at Centro Studi e Ricerche Enrico Fermi; Italy.\\
% ...

\end{flushleft}
